\documentclass[preprint]{article}       
\usepackage{graphicx}
\usepackage[english]{babel}
\usepackage[utf8]{inputenc}
\usepackage{amssymb,amsmath,amsfonts}
\usepackage{graphicx,rotating} 
\usepackage{verbatim} 
\usepackage{enumerate,graphics} 
\usepackage{multirow} 
\usepackage{url} 
\usepackage{subfigure} 
\usepackage{bm}
\usepackage{color} 
\usepackage{icomma} 
\usepackage{natbib}
\usepackage{fullpage}
\usepackage{hyperref}
\newcommand{\tablesize}{\fontsize{9}{11}\selectfont} 

\newcommand{\R}{\mathbb R} 

\allowdisplaybreaks[4] 

\begin{document}

\date{}

\title{Variable dispersion beta regressions with parametric link functions
}


\author{
Diego Ramos Canterle\thanks{Bacharelado em Estatística and LACESM, Universidade Federal de Santa Maria, Santa Maria, RS, Brazil, e-mail: \texttt{diegocanterle@gmail.com}}
\and  
F\'abio Mariano Bayer\thanks{Departamento de Estatística and LACESM, Universidade Federal de Santa Maria, Santa Maria, RS, Brazil, e-mail: \texttt{bayer@ufsm.br}}
}

\maketitle

\begin{abstract}
This paper presents a new class of regression models for continuous data restricted to the interval $(0,1)$, such as rates and proportions. The proposed class of models assumes a beta distribution for the variable of interest with regression structures for the mean and dispersion parameters. These structures consider covariates, unknown regression parameters, and parametric link functions. Link functions depend on parameters that model the relationship between the random component and the linear predictors. The symmetric and assymetric Aranda-Ordaz link functions are considered in details. Depending on the parameter values, these link functions refer to particular cases of fixed links such as logit and complementary log-log functions. Joint estimation of the regression and link function parameters is performed by maximum likelihood. Closed-form expressions for the score function and Fisher’s information matrix are presented. Aspects of large sample inferences are discussed, and some diagnostic measures are proposed. A Monte Carlo simulation study is used to evaluate the finite sample performance of point estimators. Finally, a practical application that employs real data is presented and discussed.


\end{abstract}

\section{Introduction}

The beta regression model introduced by \cite{Ferrari2004} has broad practicality for modeling variables belonging to the continuous interval $(0,1)$. In this model, it is assumed that the dependent variable $Y$ has a beta distribution, where the mean of $Y$ is modeled by a regression structure involving unknown parameters, covariates, and a link function. 
An extension of this 
model is the beta regression with varying dispersion, which has been discussed by \cite{Paolino2001}, \cite{smithson2006}, \cite{simas2010}, \cite{Pinheiro2011} and \cite{Bayer2015}. In this broader model, the dispersion parameter of $Y$ is modeled by a regression structure in the same way as the conditional mean. 
The manner in which the dispersion parameter is modeled has direct implications on the efficiency of the estimators of the mean regression structure parameters~\citep{Smyth1999, Bayer2015}. In addition to improving the inferences about the mean structure parameters, many applications are directly interested in modeling the dispersion to identify the sources of data variability~\citep{Smyth1999}.

In the variable dispersion beta regression model, the relationship between the mean and dispersion parameters of the random component $Y$ and its linear predictors are established through link functions. In this model, considering the beta density parameterization with mean $\mu \in (0,1)$ and dispersion $\sigma \in (0,1)$, as in \cite{Souza2012} and \cite{Bayer2015}, it is possible to use link functions $g(\cdot)$, such that $g(x):(0,1) \rightarrow \R$. Typical fixed link functions in these cases include the logit, probit, log-log (loglog), complementary log-log (cloglog), and Cauchy functions \citep{Koenker2009}. The fact that the possible values of $\mu$ and $\sigma$ belong to the same standard unit interval $(0,1)$ means that these link functions can be considered for both the mean and the dispersion structure.

In practice, in addition to the selection of important covariates in the mean and dispersion regression structures, as broadly discussed by \cite{Zhao2014} and \cite{Bayer2015}, the correct specification of the link functions deserves special attention. An incorrect specification of these functions may distort the inferences of the model parameters \citep[Pag. 401]{McCullagh1989} leading to misinterpretations and errors in the model predictions. To circumvent the problem of selecting an appropriate link function, a parametric link function can be considered 
\citep{Guerrero1982,Scallan1984,Stukel1988,Czado1994,Kaiser1997, Smith2003, Czado2006,Koenker2009, Adewale2010, Ramalho2011, Gomes2013, Dehbi2014, Taneichi2014, Geraci2015,  Dehbi2016}. 
Such functions involve an unknown parameter that must be estimated. In general, depending on the value of this parameter, some known link functions arise as special cases.  
The link functions proposed by Aranda-Ordaz \citep{Ordaz1981} are the parametric type most widely used in cases where the parameters of interest lie in the interval $(0,1)$. Special cases of the Aranda-Ordaz link functions include the logit and cloglog functions.

Some regression models with parametric link functions have been described in the literature. 
\cite{Guerrero1982} used a transformation of the Box-Cox link function in binary response models. 
\cite{Scallan1984} proposed generalized linear models (GLM) \citep{McCullagh1989} with general parametric link functions by presenting certain estimation aspects and identifying some special cases. \cite{Stukel1988} adjusted the binary response models to consider a two-parameter link function. \cite{Czado1994} developed a two-parameter link function that modifies the two tails of the function. 
\cite{Kaiser1997} considered the likelihood inferences of link function parameters in GLM. \cite{Czado2006} chose the link function in GLM using Bayes factors. 
\cite{Koenker2009} studied the selection of the link function in binary data using the parametric link functions of Gosset and Pregibon.
Quantile regression with Aranda-Ordaz link function is considered by \cite{Dehbi2016}.
According to \cite{Czado1997}, the maximum likelihood fit in GLM is improved by using parametric link functions in place of canonical link functions.

Regarding the beta regression model, some problems associated with the correct specification of the link function have been investigated. 
\cite{Oliveira2013} evaluated the performance of the RESET test by checking the misspecification of the link function in the beta regression model, and \cite{Pereira2013} evaluated the RESET test in the inflated beta regression model. 
\cite{Andrade2007} generalized the seminal model proposed by \cite{Ferrari2004} by considering the Aranda-Ordaz link function for the regression structure of the mean; however, this approach still considered constant dispersion.  
Nevertheless, there is a lack of studies focusing on the specification of the link function in the dispersion submodel.

Based on the above discussion, we propose a generalization of the variable dispersion beta regression model, considering parametric link functions for the  structures of both $\mu$ and $\sigma$. The parametric estimators of the link functions for the mean and dispersion submodels are proposed together with other parameters for the regression structures. The estimation of these parameters is performed using maximum likelihood estimation. Diagnostic measures and tools for model selection are also proposed.

This paper unfolds as follows. Section \ref{s:model} presents the beta regression model with parametric link functions. In Section \ref{s:emv}, we discuss  all aspects of maximum likelihood estimation. 
Section \ref{s:diag} introduces some diagnostic measures to check the goodness-of-fit in the resulting model. 
Section \ref{S:AO} presents two special cases of the parametric link functions 
based on the symmetric and asymmetric Aranda-Ordaz \citep{Ordaz1981} families of link functions. 
The finite sample performance of the estimators is assessed in Section \ref{s:simu}. 
Section \ref{s:apli} presents and discusses an application to real data on religious disbelief. Our concluding remarks are given in Section \ref{s:conclu}.

\section{The model} \label{s:model}

The beta regression model proposed by \cite{Ferrari2004} considers a constant precision parameter $\phi$ throughout the observations. Nevertheless, by erroneously assuming a constant $\phi$, the losses in efficiency for the estimators can be substantial, as discussed by \cite{Bayer2015}. In beta regression with varying dispersion, the precision parameter is assumed to be variable throughout the observations and modeled by covariates, unknown parameters, and one link function, in the same way as the mean.

In this work, as in that reported by \cite{Souza2012} and \cite{Bayer2015}, a beta density reparameterization is considered. Rather than focusing on the precision parameter $\phi$, a dispersion parameter $\sigma$ is considered. With such parameterization, the beta density is written as follows: 
\begin{align}\label{E:densidadenova}
f(y;\mu,\sigma)&=\frac{\Gamma\left(\frac{1-\sigma^2}{\sigma^2}\right)}{\Gamma\left(\mu\left(\frac{1-\sigma^2}{\sigma^2}\right)\right)\Gamma\left((1-\mu)\left(\frac{1-\sigma^2}{\sigma^2}\right)\right)}  y^{\mu\left(\frac{1-\sigma^2}{\sigma^2}\right)-1}(1-y)^{(1-\mu)\left(\frac{1-\sigma^2}{\sigma^2}\right)-1},
\end{align}
where $0<\mu<1$, $0<\sigma<1$, and $\Gamma(u)=\int_0^\infty t^{u-1} e^{-t}\rm{d}t$ is the gamma function, for $u>0$. The two parameters indexing the density assume values in the standard unit interval $(0,1)$, which enables the same link function to be used in the two regression structures. The expectation and variance of $Y$ are given by $\Bbb{E}(Y)=\mu$ and ${\rm Var}(Y)=V(\mu)\sigma^2$, respectively, where $V(\mu)=\mu(1-\mu)$ is the variance function. However, 
the proposed model is still useful for response variable restricted to the double bounded interval $(a,b)$, where $a$ and $b$ are known scalars, $a<b$. In this case, we would model $(Y-a)/(b-a)$ instead of modeling $Y$ directly \citep{Ferrari2004,smithson2006,Zimprich2010}.

Let $Y_1, \ldots, Y_n$ be independent random variables, where each $Y_t$, $t=1,\ldots, n$, has a density given by \eqref{E:densidadenova} with mean $\mu_t$ and dispersion $\sigma_t$. The variable dispersion beta regression model with parametric link functions is defined by 
\begin{align*} 
g_1(\mu_{t},\lambda_1)=\sum\limits_{i=1}^{r}x_{ti}\beta_{i}=\eta_{1t}, \\
g_2(\sigma_{t},\lambda_2)=\sum\limits_{j=1}^{s}z_{tj}\gamma_{j}=\eta_{2t},
\end{align*}
where  $\boldsymbol{\beta}=(\beta_{1}, \ldots ,\beta_{r})^{\top} \in \R^{r}$ and $
\boldsymbol{\gamma}=(\gamma_{1}, \ldots ,\gamma_{s})^{\top}\in \R^{s}$ are the vectors of unknown regression parameters ($r+s+2=q<n$), 
$\boldsymbol{x}^{\top}_{t}=(x_{t1}, \ldots, x_{tr})$ and $\boldsymbol{z}^{\top}_{t}=(z_{t1}, \ldots, z_{ts})$ represent the $t$th observations of the explanatory variables, which are assumed to be fixed and known, 
and 
$\eta_{1t}=\boldsymbol{x}^{\top}_{t}\boldsymbol{\beta}$ and $\eta_{2t}=\boldsymbol{z}^{\top}_{t}\boldsymbol{\gamma}$ are the linear predictors for the mean and dispersion, respectively. 
Finally, $g_1(\cdot,\cdot)$ and $g_2(\cdot,\cdot)$ are strictly monotonic in the first argument and twice differentiable in both arguments, such that $g_\delta :(0,1)\rightarrow\R$, for $\delta=1,2$. 
The second arguments of $g_\delta(\cdot,\cdot)$, 
$\lambda_1\in \Lambda_1$ and $\lambda_2\in \Lambda_2$, 
are the link function parameters.
Further, note that 
\begin{align}
\mu_t=g^{-1}_1(\eta_{1t},\lambda_1),\label{E:invmu}\\ 
\sigma_t=g^{-1}_2(\eta_{2t},\lambda_2).\label{E:invsigma}
\end{align}
The parameters $\lambda_1$ and $\lambda_2$ are shape parameters that generally influence the symmetry and heaviness of tails of the fitted curves for $\mu$ and $\sigma$ \citep{Stukel1988}.

Unlike models that consider fixed link functions, the proposed model captures different relationships between the linear predictors $\eta_{\delta t}$, $\delta=1,2$, and their respective parameters $\mu_t$ and $\sigma_t$. Depending on the parametric value $\lambda$ for a given function $g(\cdot, \lambda )$, there is a particular family of link functions given by 
\begin{align*}
\mathcal{G} = \left\lbrace g(\cdot,\lambda): \lambda \in \Lambda \right\rbrace.
\end{align*}
Different link function families can be considered. When the parameters of interest are in the continuous interval $(0,1)$, such as $\mu_t$ and $\sigma_t$ in the proposed model, possibilities include the symmetric and asymmetric link functions proposed by \cite{Ordaz1981}, 
Box-Cox transformation link function~\citep{Guerrero1982}, 
Gosset link function~\citep{Koenker2009}, 
Pregibon link function~\citep{Pregibon1980}, 
and generalized logit function considered by \cite{Ramalho2011}. 
In particular, the Pregibon link function has two parameters, and is not contextualized in this work. 
Gosset link function fails to consider
the possible asymmetric relationship between the random component and the linear predictors. 
In this regard, and in addition to the overall results of any one-parametric links, 
this work presents results for  
the
symmetric and 
asymmetric 
Aranda-Ordaz link functions.

\section{Likelihood inference} \label{s:emv}

The maximum likelihood estimation of the parametric vector $\boldsymbol{\theta}=(\boldsymbol{\beta}^\top \!, \boldsymbol{\gamma}^\top \!, \lambda_1, \lambda_2)^\top$ is given by maximizing the logarithm of the likelihood function. Given a sample size $n$ and considering the form of the density in \eqref{E:densidadenova}, the log-likelihood is given by 
\begin{align}\label{E:logvero} 
\ell(\boldsymbol{\theta})=\sum\limits_{t=1}^{n} \ell_{t}(\mu_{t},\sigma_{t}),
\end{align}
where 
\begin{align*}
\ell_{t}(\mu_{t},\sigma_t)&=\log\Gamma\left(\frac{1-\sigma_t^2}{\sigma_t^2}\right)-\log\Gamma\left(\mu_{t}\frac{1-\sigma_t^2}{\sigma_t^2}\right)-\log \Gamma\left((1 - \mu_{t})\frac{1-\sigma_t^2}{\sigma_t^2}\right) \\
&+   \left(\mu_{t}\frac{1-\sigma_t^2}{\sigma_t^2}-1\right) \log y_{t}+\left((1-\mu_{t})\frac{1-\sigma_t^2}{\sigma_t^2}-1\right)\log(1-y_{t}),
\end{align*} 
in which $\mu_t$ and $\sigma_t$ are given by the regression structures in \eqref{E:invmu} and \eqref{E:invsigma}, respectively.

By deriving the log-likelihood function in \eqref{E:logvero} with respect to the parametric vector $\boldsymbol{\theta}$, we obtain the score vector 
$U(\boldsymbol{\theta})= 
\left( 
U_{\boldsymbol{\beta}}(\boldsymbol{\theta})^\top, 
U_{\boldsymbol{\gamma}}(\boldsymbol{\theta})^\top,
U_{\lambda_1}(\boldsymbol{\theta}),
U_{\lambda_2}(\boldsymbol{\theta})
\right)^\top
$. 
Details of the analytical derivations are given in detail in the Appendix. The score function with respect to $\beta$ is given by 
\begin{align*}
U_{\boldsymbol{\beta}}(\boldsymbol{\theta})= \boldsymbol{X}^{\top}\boldsymbol{\Sigma} \boldsymbol{T}(\boldsymbol{y}^*-\boldsymbol{\mu}^*),
\end{align*}
where 
$\boldsymbol{X}$ is the $n\times r$ matrix in which the $t$th row is $\boldsymbol{x}_t$, $\boldsymbol{\Sigma}\!=\!{\rm diag}\!\left(\!\frac{1-\sigma_1^2}{\sigma_1^2}, \ldots ,\! \frac{1-\sigma_n^2}{\sigma_n^2}\!\right)$, $\boldsymbol{T} = {\rm diag} \bigg( \left[\frac{\partial g_1(\mu_{1},\lambda_1)}{\partial\mu_1}\right]^{-1},$
$ \ldots , \left[\frac{\partial g_1(\mu_{n},\lambda_1)}{\partial\mu_n}\right]^{-1}\bigg)$, $\boldsymbol{y}^*=(y^*_1,\ldots,y^*_n)^{\top}$, $\boldsymbol{\mu}^*=(\mu^*_1,\ldots,\mu^*_n)^{\top}$, with $y_t^*=\log (y_t/(1-y_t))$, $\mu_t^*=\psi\left(\mu_t\frac{1-\sigma_t^2}{\sigma_t^2}\right)-\psi\left((1-\mu_t)\frac{1-\sigma_t^2}{\sigma_t^2}\right)$, 
and $\psi(\cdot)$ is the digamma function, i.e., $\psi(u)=\frac{d \log \Gamma(u)}{d u}$.

The score function with respect to $\boldsymbol{\gamma}$ is given by 
\begin{align*}
U_{\boldsymbol{\gamma}}(\boldsymbol{\theta})= \boldsymbol{Z}^{\top}\boldsymbol{H}\boldsymbol{a},
\end{align*}
where $\boldsymbol{Z}$ is the $n\times s$ matrix whose $t$th row is $\boldsymbol{z}_t$, 
$\boldsymbol{H}= {\rm diag} \left( \left[\frac{\partial g_2(\sigma_{1},\lambda_2)}{\partial\sigma_1}\right]^{-1},\ldots ,  \left[\frac{\partial g_2(\sigma_{n},\lambda_2)}{\partial\sigma_n}\right]^{-1}\right)$, 
$\boldsymbol{a}=(a_1,\ldots,a_n)^{\top}$, with 
\begin{align*}
a_t=-\dfrac{2}{\sigma^3_t}\bigg[\mu_t(y_t^*-\mu_t^*)+\psi\left(\frac{1-\sigma_t^2}{\sigma_t^2}\right) -\psi\left((1-\mu_t)\frac{1-\sigma_t^2}{\sigma_t^2}\right)+\log (1-y_t)\bigg].
\end{align*}

The score functions with respect to $\lambda_1$ and $\lambda_2$ are given by 
\begin{align*}
U_{\lambda_1}(\boldsymbol{\theta})&=\sum_{t=1}^{n}\frac{1-\sigma_t^2}{\sigma_t^2}(y^*_t-\mu^*_t)\rho_t,\\
U_{\lambda_2}(\boldsymbol{\theta})&=\sum_{t=1}^{n}a_t\varrho_t,
\end{align*}
respectively, where $\rho_t = \dfrac{\partial \mu_t}{\partial \lambda_1}$ depends on the parametric link function to be used in the mean submodel 
and 
$\varrho_t = \dfrac{\partial\sigma_t}{\partial \lambda_2}$ depends on the link function considered in the dispersion submodel.
In Section \ref{S:AO}, the quantities $\rho_t = \dfrac{\partial \mu_t}{\partial \lambda_1}$ and $\varrho_t = \dfrac{\partial\sigma_t}{\partial \lambda_2}$ are presented for the 
symmetric and
asymmetric Aranda-Ordaz link functions.

The maximum likelihood estimators (MLEs) for the beta regression model with parametric link functions are obtained by solving the following nonlinear system: 
\begin{align}\label{E:vescore}
\left\{ \begin{array}{ll} 
U_{\boldsymbol{\beta}}(\boldsymbol{\theta})&= 0 \\
U_{\boldsymbol{\gamma}}(\boldsymbol{\theta})&= 0 \\
U_{\lambda_1}(\boldsymbol{\theta})&= 0 \\
U_{\lambda_2}(\boldsymbol{\theta})&= 0
\end{array} \right. .
\end{align}
Solving Equation \eqref{E:vescore} requires the use of nonlinear optimization algorithms. In this work, the quasi-Newton BFGS method~\citep{press} was used for the computational implementations.

Fisher's information matrix, which is useful for large sample inferences, requires the expectations of the second derivatives of the log-likelihood function. 
Details of the analytical derivation of these quantities are given in the Appendix. The joint information matrix for the parametric vector $\boldsymbol{\theta}$ is given by 
\begin{align}\label{E:MIF}
\boldsymbol{K}=K(\boldsymbol{\theta})=
\begin{pmatrix}
K_{(\beta,\beta)} &  K_{(\beta, \gamma)} &  K_{(\beta,\lambda_1)} &  K_{(\beta, \lambda_2)} \\
K_{(\gamma,\beta)} &  K_{(\gamma, \gamma)} & K_{(\gamma,\lambda_1)} &  K_{(\gamma, \lambda_2)} \\
K_{(\lambda_1,\beta)} &  K_{(\lambda_1, \gamma)} &  K_{(\lambda_1,\lambda_1)} &  K_{(\lambda_1, \lambda_2)} \\
K_{(\lambda_2,\beta)} &  K_{(\lambda_2, \gamma)} & K_{(\lambda_2,\lambda_1)}  & K_{(\lambda_2, \lambda_2)} 
\end{pmatrix},
\end{align} 
where $K_{(\beta ,\beta)}= \boldsymbol{X}^\top \boldsymbol{\Sigma} \boldsymbol{W} \boldsymbol{X}$, $K_{(\beta,\gamma)}=K_{(\gamma ,\beta )}^{\top }=\boldsymbol{X}^{\top }\boldsymbol{C} \boldsymbol{T} \boldsymbol{H} \boldsymbol{Z}$, $K_{(\beta,\lambda_1)}=K_{(\lambda_1,\beta )}^{\top }=\boldsymbol{X}^{\top }\boldsymbol{V} \boldsymbol{T} \boldsymbol{\rho}$, $K_{(\beta,\lambda_2)}=K_{(\lambda_2,\beta )}^{\top }=\boldsymbol{X}^{\top }\boldsymbol{C} \boldsymbol{T} \boldsymbol{\varrho}$, $K_{(\gamma,\gamma)} = \boldsymbol{Z}^{\top}\boldsymbol{D}^*\boldsymbol{H}\boldsymbol{H}^{\top}\boldsymbol{Z}$, $K_{(\gamma,\lambda_1)} = K_{(\lambda_1,\gamma)}^{\top} = \boldsymbol{Z}^{\top}\boldsymbol{C}\boldsymbol{H}\boldsymbol{\rho}$, $K_{(\gamma,\lambda_2)} = K_{(\lambda_2,\gamma)}^{\top} = \boldsymbol{Z}^{\top}\boldsymbol{D}^*\boldsymbol{H}\boldsymbol{\varrho}$, $K_{(\lambda_1,\lambda_1)} = \boldsymbol{\rho}^{\top}\boldsymbol{V}\boldsymbol{\rho}$, $K_{(\lambda_1,\lambda_2)} = K_{(\lambda_2,\lambda_1)}^{\top} = \boldsymbol{\rho}^{\top}\boldsymbol{C}\boldsymbol{\varrho}$, and $K_{(\lambda_2,\lambda_2)} = \boldsymbol{\varrho}^{\top}\boldsymbol{D}^*\boldsymbol{\varrho}$, with $\boldsymbol{\rho}=(\rho_1,\ldots,\rho_n)^{\top}$, $\boldsymbol{\varrho}=(\varrho_1,\ldots,\varrho_n)^{\top}$, $\boldsymbol{W} = {\rm diag}(w_1,\ldots,w_n)$, $\boldsymbol{C} = {\rm diag}(c_1,\ldots,c_n)$, $\boldsymbol{V} = {\rm diag}(\nu_1,\ldots,\nu_n)$, and $\boldsymbol{D}^* = {\rm diag}(d^*_1,\ldots,$\\$d^*_n)$. 
Finally,
\begin{align*}
w_t &= \dfrac{1-\sigma_t^2}{\sigma_t^2} \left[ \psi'\left( \mu_t \dfrac{1-\sigma_t^2}{\sigma_t^2}  \right) + \psi'\left( (1-\mu_t) \dfrac{1-\sigma_t^2}{\sigma_t^2}  \right)  \right] \left( \dfrac{\partial g_1(\mu_t,\lambda_1)}{\partial\mu_t} \right)^{-2},\\
c_t &= \dfrac{1-\sigma^2_t}{\sigma^2_t}\dfrac{2}{\sigma_t^3}\bigg[ (1-\mu_t)\psi'\left( (1-\mu_t) \dfrac{1-\sigma_t^2}{\sigma_t^2}  \right) -\mu_t\psi'\left( \mu_t \dfrac{1-\sigma_t^2}{\sigma_t^2}  \right)   \bigg],\\
\nu_t &= \left(\dfrac{1-\sigma_t^2}{\sigma_t^2}\right)^2 \bigg[ \psi'\left( \mu_t \dfrac{1-\sigma_t^2}{\sigma_t^2}  \right) + \psi'\left( (1-\mu_t) \dfrac{1-\sigma_t^2}{\sigma_t^2}  \right)  \bigg],\\
d_t^* &= \dfrac{4}{\sigma^6_t}\bigg[\!-\!\psi'\left(\dfrac{1\!-\!\sigma_t^2}{\sigma_t^2}\right) + \mu_t^2\psi'\left(\mu_t\dfrac{1\!-\!\sigma_t^2}{\sigma_t^2}\right) 
+ (1\!-\!\mu_t)^2\psi'\left((1\!-\!\mu_t)\dfrac{1\!-\!\sigma_t^2}{\sigma_t^2}\right)\bigg],
\end{align*}
where $\psi'(\cdot)$ is the trigamma function, i.e., $\psi'(u)=\frac{d \psi(u)}{d u}$, for $u>0$. According to the concept of orthogonality by \cite{Cox1987}, \eqref{E:MIF} can be used to ascertain that the model parameters are not orthogonal because the information matrix is not a diagonal block matrix.

\subsection{Large sample inference} \label{S:IGA}

Under the usual regularity conditions for MLE \citep{Pawitan2001}, 
the joint distribution of the MLEs is approximately $q$-multivariate normal when the sample size is large, i.e.,
\begin{align*}
\left(\begin{array}{llll}
\widehat{\boldsymbol{\beta}}\\
\widehat{\boldsymbol{\gamma}}\\
\widehat{\lambda}_1\\
\widehat{\lambda}_2
\end{array} \right ) \sim \mathcal{N}_{q}
\left(\begin{array}{llll}
\left(\begin{array}{llll}
\boldsymbol{\beta} \\
\boldsymbol{\gamma} \\
\lambda_1\\
\lambda_2
\end{array} \right ), \boldsymbol{K}^{-1}
\end{array} \right ) ,
\end{align*}
where $\widehat{\boldsymbol{\beta}}$, $\widehat{\boldsymbol{\gamma}}$, $\widehat{\lambda}_1$, and $\widehat{\lambda}_2$ are the MLEs of $\boldsymbol{\beta}$, $\boldsymbol{\gamma}$, $\lambda_1$, and $\lambda_2$, respectively, and $\boldsymbol{K}^{-1}$ is the inverse Fisher's information matrix.

The Wald confidence intervals model parameters $\theta_m$, $m=1,\ldots,q$, are defined by \citep{Pawitan2001,Ferrari2004}:
\begin{align*}
[\widehat{\theta}_m - \Phi^{-1}(1-\alpha/2)\widehat{{\rm se}}(\widehat{\theta}_m) ; \widehat{\theta}_m + \Phi^{-1}(1-\alpha/2)\widehat{{\rm se}}(\widehat{\theta}_m)],
\end{align*}
where $\widehat{\theta}_m$ represents the MLE of $\theta_m$, 
the standard error of $\widehat{\theta}_m$ is given by $\widehat{{\rm se}}(\widehat{\theta}_m)=[{\rm diag}({\rm \widehat{cov}}(\widehat{\boldsymbol{\theta}}))]_m^{1/2}$, 
in which ${\rm \widehat{cov}}(\widehat{\boldsymbol{\theta}}) = \boldsymbol{K}^{-1}(\widehat{\boldsymbol{\theta}})$ is the asymptotic variance and covariance matrix of $\widehat{\boldsymbol{\theta}}$, 
$\Phi^{-1}$ is the quantile function of the standard normal distribution, 
and $\alpha$ is the nominal level of the confidence interval. 
Similar to \cite{Ferrari2004}, for $\mu_t$ and $\sigma_t$, for $\delta=1,2$ respectively, we have the following confidence intervals:
\begin{align*}
[g_\delta^{-1}(\widehat{\eta}_{\delta t}-\Phi^{-1}(1-\alpha/2)\widehat{{\rm se}}(\widehat{\eta}_{\delta t}),\widehat{\lambda}_\delta); g_\delta^{-1}(\widehat{\eta}_{\delta t}+\Phi^{-1}(1-\alpha/2)\widehat{{\rm se}}(\widehat{\eta}_{\delta t}),\widehat{\lambda}_\delta)],
\end{align*}
where the standard errors of $\widehat{\eta}_{\delta t}$, for $\delta=1,2$, are estimated by $\widehat{{\rm se}}(\widehat{\eta}_{1t})=(x_t \widehat{{\rm cov}}(\widehat{\beta}) x_t^\top)^{1/2}$  and $\widehat{{\rm se}}(\widehat{\eta}_{2t})=(z_t \widehat{{\rm cov}}(\widehat{\gamma}) z_t^\top)^{1/2}$.

To test the hypotheses on the parameters, we consider the null hypothesis $\mathcal{H}_0:\theta_m=\theta_{m}^0$ versus $\mathcal{H}_1:\theta_m \neq \theta_{m}^0$. The Wald test can be considered by using the following statistic \citep{Pawitan2001}: 
\begin{align*}
z = \dfrac{\widehat{\theta}_m-\theta_{m}^0}{\widehat{{\rm se}}(\widehat{\theta}_m)}.
\end{align*}
Because the $z$ statistic has an asymptotically standard normal distribution under $\mathcal{H}_0$, the test is performed by comparing the calculated $z$ statistic with the usual quantiles of the standard normal distribution.

For more general hypotheses, $\mathcal{H}_0:\boldsymbol{\theta}_I=\boldsymbol{\theta}_I^0$ versus $\mathcal{H}_1:\boldsymbol{\theta}_I\neq\boldsymbol{\theta}_I^0$, where $\boldsymbol{\theta}=(\boldsymbol{\theta}_I^\top,\boldsymbol{\theta}_N^\top)^\top$ has dimension $q$, $\boldsymbol{\theta}_I$ is the vector of parameters of interest with dimension $\iota$, and $\boldsymbol{\theta}_N$ is the vector of nuisance parameters with dimension $q-\iota$, four test statistics can be considered, namely: the likelihood ratio (LR)~\citep{Person1928}, Wald (W)~\citep{Wald1943}, score (S)~\citep{Rao1948}, and gradient (G)~\citep{Terrell2002}. 
Under $\mathcal{H}_0$ and the usual conditions of regularity, the four test statistics have the asymptotic chi-squared distribution with $\iota$ degrees of freedom ($\chi_{\iota}^{2}$), where $\iota$ is the number of restrictions imposed by the null hypothesis \citep{Vargas2014}. The test can be performed by comparing the calculated value of the statistic considered, i.e., LR, W, S, or G, with the usual quantile of $\chi_{\iota}^{2}$.

\section{Diagnostics} \label{s:diag}

After estimating the model, it is necessary to evaluate possible departures from the model assumptions, as well as the detection of unadjusted or aberrant points. This section introduces some diagnostic measures to determine the correct adjustment of the proposed model.

Residuals are an important measure in checking for deviations from the unknown population model, disparate observations, and adjustment quality. Initially, the standardized ordinary residual is proposed. This is given by 
\begin{align*}
r_t = \dfrac{y_t-\widehat{\mu}_t}{\sqrt{\widehat{\rm Var}(Y_t)}},
\end{align*}
where ${\rm \widehat{Var}}(Y_t)=\widehat{\mu}_t(1-\widehat{\mu}_t)\widehat{\sigma}_t^2$. Additionally, the standardized weighted residual 2 can be used, as proposed by \cite{Ferrari2011} for the varying dispersion beta regression model. This is given by 
\begin{align*}
r_t^{pp} = \dfrac{y_t^*-\widehat{\mu}_t^*}{\sqrt{{\rm \widehat{Var}}(y_t^*)(1-h_{tt})}},
\end{align*}
where ${\rm \widehat{Var}}(y_t^*) = \psi'\left(\widehat{\mu}_t\frac{1-\widehat{\sigma}_t^2}{\widehat{\sigma}_t^2}\right)-\psi'\left((1-\widehat{\mu}_t)\frac{1-\widehat{\sigma}_t^2}{\widehat{\sigma}_t^2}\right)$, and $h_{tt}$ is the $t$th diagonal element of the `hat matrix' $\mathbf{H}=(\widehat{\boldsymbol{W}}\widehat{\boldsymbol{\Sigma}})^{1/2}\boldsymbol{X}(\boldsymbol{X}^\top \widehat{\boldsymbol{\Sigma}}\widehat{\boldsymbol{W}}\boldsymbol{X})^{-1}\boldsymbol{X}^\top (\widehat{\boldsymbol{\Sigma}}\widehat{\boldsymbol{W}})^{1/2}$. 
This residual provides an improved approximation of the standard normal distribution when the model is correctly adjusted and when a model with fixed links is considered \citep{espinheira2008b}. In prior simulations and analyses, the performance of the $r_t^{pp}$ residuals was found to be good in the proposed model considering parametric links. A residual chart is typically used to analyze the residuals against their respective indices. In this chart, the residuals are expected to be randomly distributed around zero, and no more than $5\%$ of the 
values can occur outside of the $[-2,2]$ interval.

To verify that the distribution assumed for the dependent variable is adequate, we can examine half-normal plots with simulated envelopes by evaluating the quality of the fitted model \citep{Atkinson1981}. The simulated envelope can be built as follows \citep{Atkinson1985,Ferrari2004}: 
\begin{enumerate}[(i)]
\item \label{i} fit the model and generate a simulated sample set of $n$ independent observations using the fitted model as if it were the true model;

\item \label{ii} fit the model from the generated sample, calculate the absolute values of the residuals and arrange them in order;

\item repeat steps  \eqref{i} and  \eqref{ii} $k$ times;

\item consider the $n$ sets of the $k$ order statistics; for each set, calculate the quantile  $\alpha/2$, the mean, and the quantile $1-\alpha/2$;

\item plot these values and the ordered residuals of the original sample set against the $\Phi^{-1}((t+n+1/2)/(2n+10/8))$ scores. 
\end{enumerate}
No more than $\alpha\times 100\%$ of the observations are expected to occur outside the envelope bands. A very large proportion of points lying outside the bands suggests that the model is inadequate.

The overall influence measures of each observation under the estimates of the model parameters can be considered using Cook's distance \citep{Cook1977}. In this study, we use the Cook-like distance proposed by \cite{espinheira2008} for the beta regression model. This distance combines leverage measures and the model residuals, and is defined by 
\begin{align*}
C_t = \dfrac{h_{tt}}{1-h_{tt}}(r_t^{pp})^2.
\end{align*}
To check for possible points of influence, it is common to produce a chart of $C_t$ against their respective $t$ indices.

Candidate models can be selected using information criteria, such as the generalized Akaike information criterion (GAIC)~\citep{Akaike1983,Rigby2005}, which is given by 
\begin{align*}
{\rm GAIC} = -2\ell(\widehat{\boldsymbol{\theta}})+\mathcal{P}q,
\end{align*} 
where $\mathcal{P}$ can take different real values. Values of $\mathcal{P}=2$ and $\mathcal{P}={\rm log}(n)$, give the Akaike information criterion (AIC) \citep{Akaike1974} and the Schwarz information criterion (SIC) \citep{Schwarz1978}, respectively. These criteria take into account the maximized log-likelihood penalized by the number of parameters in the adjusted model. For the selection of competitive models, that with the lowest GAIC value should be chosen.
 
To ascertain the correct model specification, the RESET tests \citep{Ramsey1969} are recommended. \cite{McCullagh1989} suggested using a RESET-type test in GLM, whereas \cite{Pereira2013} and \cite{Oliveira2013} argued they are suitable for the beta regressions. To run the RESET-type test for the proposed model, $\widehat{\boldsymbol{\eta}}_1^2$ should be added as a covariate in both the mean and dispersion submodels.
This new model should be fitted with $\lambda_1$ and $\lambda_2$ fixed to their previously estimated values. The parameters of the artificial covariates $\widehat{\boldsymbol{\eta}}_1^2$ should then be tested according to the $\mathcal{H}_0:(\boldsymbol{\beta}_{r+1},\boldsymbol{\gamma}_{s+1})=(0,0)$ null hypothesis, where $\boldsymbol{\beta}_{r+1}$ and $\boldsymbol{\gamma}_{s+1}$ are the parameters pertaining to the artificial covariates in the mean and dispersion submodels,
respectively. If $\mathcal{H}_0$ is not rejected, the model is specified correctly; otherwise, the model is specified incorrectly. To run the RESET-type test, any one of the four test statistics cited in Subsection \ref{S:IGA} can be used.

We can use the LR, W, S, and G statistics 
to test the incorrect specification of some fixed link function. 
Considering the asymmetric Aranda-Ordaz link function, 
we can test $H_0: (\lambda_1, \lambda_2) = (1,1)$ to check whether the logit link function for mean and dispersion submodels is appropriate. 
If $\mathcal{H}_0$ is not rejected, the fixed logit links are correctly specified.

As a global measure of the goodness-of-fit, we consider the generalized coefficient of determination \citep{Nagelkerke1991}. This is given by 
\begin{align*}
R^2_{G} = 1 - \left( \dfrac{L_{null}}{L_{fit}} \right)^{(2/n)} = 1 - {\rm exp}\left( -\dfrac{2}{n}\left[ \ell(\widehat{\boldsymbol{\theta}})-\ell(0) \right] \right),
\end{align*} 
where $\ell(0)$ is the maximized log-likelihood of the null model, i.e., under constant mean and dispersion\footnote{When constant mean and dispersion are considered, no regression structures are considered; thus, there are no estimates for $\lambda_\delta$.}, $\ell(\widehat{\boldsymbol{\theta}})$ is the maximized log-likelihood of the fitted model, $\ell(0)={\rm log}L_{null}$, and $\ell(\widehat{\boldsymbol{\theta}})={\rm log}L_{fit}$. $R^2_{G}$ measures the proportion of the variability of $Y$ that can be explained by the fitted  model; this lies in the interval $[0,1]$. A higher value of $R^2_{G}$ implies that the model predictions are more accurate.

\section{Aranda-Ordaz link functions - two particular cases} \label{S:AO}

As mentioned earlier in this paper, 
the Aranda-Ordaz link function families \citep{Ordaz1981} 
can be used to relate the mean and dispersion parameters with their respective linear predictors. 
We considered 
these link functions because they 
are two one-parameter families of symmetric and asymmetric links that 
includes several well-known links as particular cases \citep{Dehbi2016}.  
They can be also considered in several works 
in a multitude of regression models 
\citep{Morgan1992, Colosimo2000, Smith2003, Adewale2010, Gomes2013, Dehbi2014, Taneichi2014, Geraci2015,  Dehbi2016}. 
Because the two parameters $\mu$ and $\sigma$ of the proposed model assume values in the same interval $(0,1)$, the relationships established immediately below are valid for both of these parameters.

\begin{figure}[t]
\begin{center}
\subfigure[Symmetric Aranda-Ordaz]{
{\includegraphics[width=0.47\textwidth]{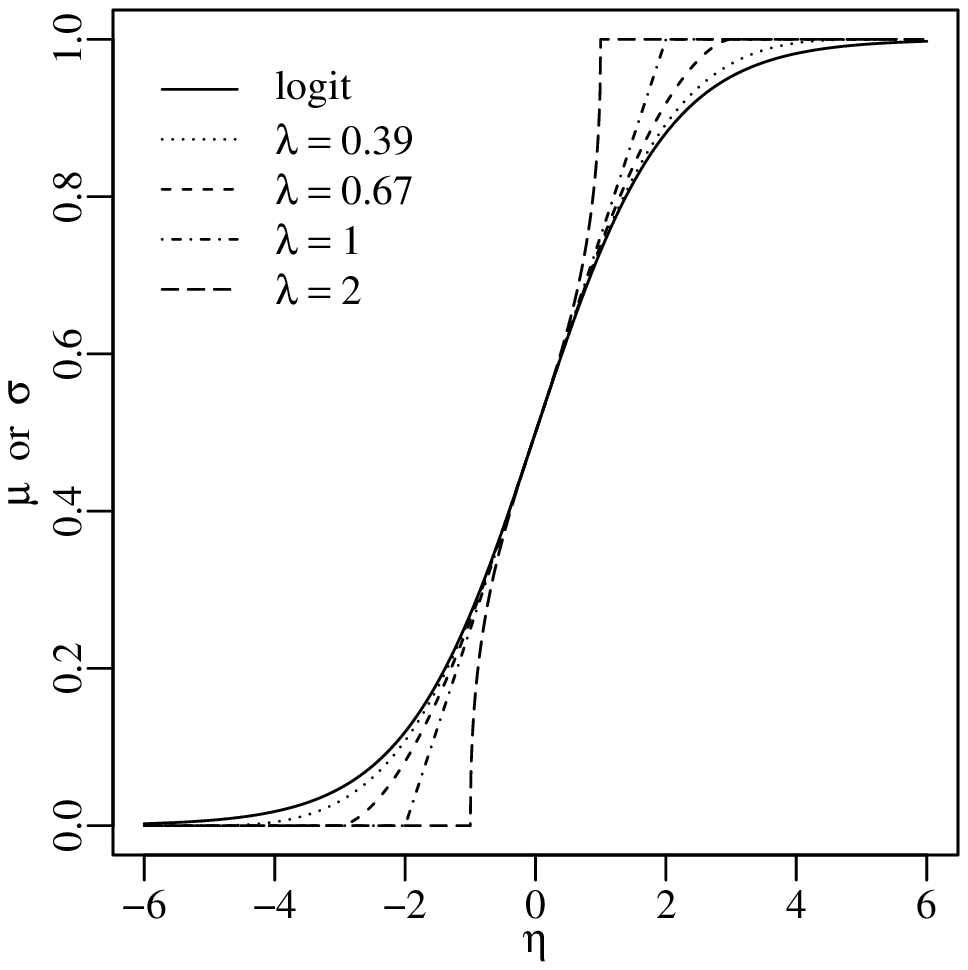}}\label{F:ao-sym}}
\quad 
\subfigure[Asymmetric Aranda-Ordaz]{
{\includegraphics[width=0.47\textwidth]{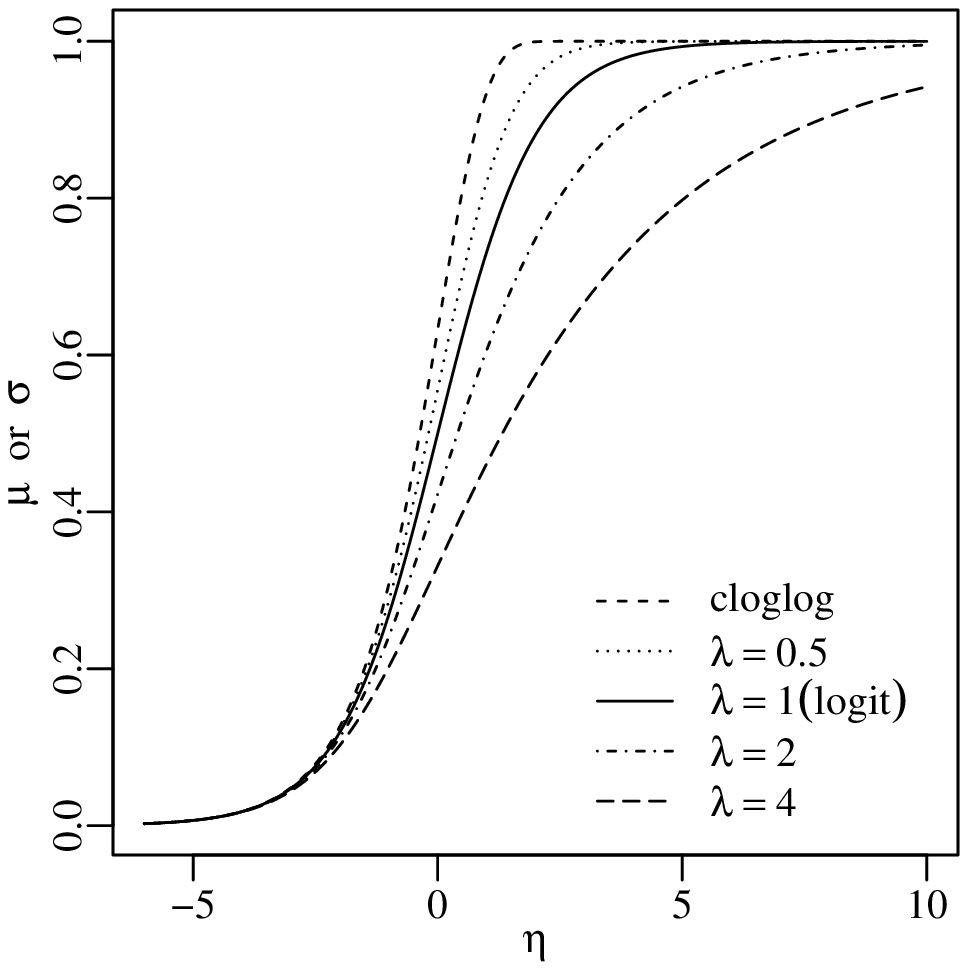}}\label{F:ao-asym}}
\caption{Aranda-Ordaz link functions for different values of $\lambda$.}
\label{F:f1}
\end{center}
\end{figure}

The symmetric Aranda-Ordaz link function is given by:
\begin{align*}
\eta = {\rm g}(\mu,\lambda) = \frac{2 \left(\mu^{\lambda }-(1-\mu)^{\lambda }\right)}{\lambda  \left(\mu^{\lambda }+(1-\mu)^{\lambda }\right)},
\end{align*}
where $\lambda \neq 0$ 
and $\mu \in (0,1)$. 
The symmetry refers to the fact that 
${\rm g}(\mu,\lambda) = -{\rm g}(1-\mu,\lambda)$ 
and 
${\rm g}(\mu,\lambda) = {\rm g}(\mu,-\lambda)$ \citep{Dehbi2016}. 
This link function family reduces 
to the linear link function if $\lambda = 1$, 
to the logit if $\lambda \rightarrow 0$, 
close to the probit link if $\lambda = 0.39$, 
and 
close to the arc sine link function if $\lambda = 0.67$ \citep{Ordaz1981, Dehbi2016}. 
Figure~\ref{F:ao-sym} shows 
some different forms of the symmetric Aranda-Ordaz link function 
considering different values of the link function parameter $\lambda$. 
For this symmetric link function, the inverse function 
can be written as follows: 
\begin{align*}
\mu = {\rm g}^{-1}(\eta,\lambda) = \frac{\left(\frac{\lambda  \eta}{2}+1\right)^{\frac{1}{\lambda }}}{\left(1-\frac{\lambda  \eta}{2}\right)^{\frac{1}{\lambda }}+\left(\frac{\lambda 
\eta}{2}+1\right)^{\frac{1}{\lambda }}}.
\end{align*}

In the general formulation of the proposed model presented in Section \ref{s:emv}, 
the score vector and Fisher's information matrix involve the quantities 
$\left(\frac{\partial g_1(\mu_t,\lambda_1)}{\partial\mu_t}\right)^{-1}$, 
$\left(\frac{\partial g_2(\sigma_t,\lambda_2)}{\partial\sigma_t}\right)^{-1}$, 
$\boldsymbol{\rho}$, and $\boldsymbol{\varrho}$, 
which depend on the considered parametric link functions. 
Considering the symmetric Aranda-Ordaz link function in both regression structures, we have: 
\begin{align*}
&\dfrac{\partial g_1(\mu_t,\lambda_1)}{\partial\mu_t} = 
\frac{4 (\mu_t (1-\mu_t))^{\lambda_1 -1}}{\left(\mu_t^{\lambda_1 }+(1-\mu_t)^{\lambda_1 }\right)^2},\\
&\dfrac{\partial g_2(\sigma_t,\lambda_2)}{\partial\sigma_t} = 
\frac{4 (\sigma_t (1-\sigma_t))^{\lambda_2 -1}}{\left(\sigma_t^{\lambda_2 }+(1-\sigma_t)^{\lambda_2 }\right)^2},\\
&\dfrac{\partial \mu_t}{\partial \eta_{1t}} =\frac{4 \left(4-\lambda_1^2 \eta_{1t}^2\right)^{\frac{1}{\lambda_1}-1}}{\left((2-\lambda_1  \eta_{1t})^{\frac{1}{\lambda_1}}+(\lambda_1  \eta_{1t}+2)^{\frac{1}{\lambda_1}}\right)^2},\\
&\dfrac{\partial \sigma_t}{\partial \eta_{2t}} =\frac{4 \left(4-\lambda_2^2 \eta_{2t}^2\right)^{\frac{1}{\lambda_2}-1}}{\left((2-\lambda_2  \eta_{2t})^{\frac{1}{\lambda_2}}+(\lambda_2  \eta_{2t}+2)^{\frac{1}{\lambda_2}}\right)^2},\\
&\rho_t =\dfrac{\partial\mu_t}{\partial\lambda_1}  
= \frac{2 \left(4-\lambda_1 ^2 \eta_{1t}^2\right)^{\frac{1}{\lambda_1 }-1} \left(\left(\lambda_1 ^2 \eta_{1t}^2-4\right) \tanh ^{-1}\left(\frac{\lambda_1  \eta_{1t}}{2}\right)+2 \lambda_1 
   \eta_{1t}\right)}{\lambda_1 ^2 \left((2-\lambda_1  \eta_{1t})^{\frac{1}{\lambda_1 }}+(\lambda_1  \eta_{1t}+2)^{\frac{1}{\lambda_1 }}\right)^2},
   \end{align*}
and 
\begin{align*}
\varrho_t&=
\dfrac{\partial\mu_t}{\partial\lambda_2}  
= \frac{2 \left(4-\lambda_2 ^2 \eta_{2t}^2\right)^{\frac{1}{\lambda_2 }-1} \left(\left(\lambda_2 ^2 \eta_{2t}^2-4\right) \tanh ^{-1}\left(\frac{\lambda_2  \eta_{2t}}{2}\right)+2 \lambda_2 
   \eta_{2t}\right)}{\lambda_2 ^2 \left((2-\lambda_2  \eta_{2t})^{\frac{1}{\lambda_2 }}+(\lambda_2  \eta_{2t}+2)^{\frac{1}{\lambda_2 }}\right)^2}. 
\end{align*}

The asymmetric Aranda-Ordaz link function is given by~\citep{Ordaz1981}:
\begin{align*}
\eta = g(\mu,\lambda) = {\rm log} \left(  \dfrac{(1-\mu)^{-\lambda}-1}{\lambda} \right) ,
\end{align*}
where $\lambda > -1/e^{\eta}$, $\mu \in (0,1)$, 
and its inverse can be written as follows: 
\begin{align*}
\mu = g^{-1}(\eta,\lambda) = 1-\left[ 1+\lambda {\rm exp}(\eta) \right]^{-\frac{1}{\lambda}}. 
\end{align*}
The asymmetric Aranda-Ordaz function 
is more flexible than the symmetric version 
and it captures the possible asymmetry between the linear predictors and the parameters $\mu$ and $\sigma$. 
In Figure~\ref{F:ao-asym}, this relationship can be seen for different values of the 
 parameter $\lambda$. 
 The logit and cloglog link functions are special cases 
for $\lambda = 1$ and $\lambda \rightarrow 0$, respectively. 
Compared with the usual logit function, $\mu$ or $\sigma$  tends to 1 more quickly as $\eta_\delta$ increases when $\lambda<1$; 
and for $\lambda>1$, the parameters $\mu$ or $\sigma$ tends more slowly to 1 as $\eta_\delta$ increases.  
It is notable that a link function with a lower parameter value results in a greater variation in $\mu$ and/or $\sigma$ in relation to $\eta_\delta$. In contrast, very high values for the link function parameter might indicate that the parameters $\mu$ and/or $\sigma$ are not variable and should be estimated without independent variables, i.e., as constants.

Considering the asymmetric Aranda-Ordaz link function 
the quantities needed for score vector and Fisher's information matrix 
are given by:
\begin{align*}
&\dfrac{\partial g_1(\mu_t,\lambda_1)}{\partial\mu_t}  = \dfrac{\lambda_1(1-\mu_t)^{-(\lambda_1+1)}}{(1-\mu_t)^{-\lambda_1}-1},\\
&\dfrac{\partial g_2(\sigma_t,\lambda_2)}{\partial\sigma_t}  = \dfrac{\lambda_2(1-\sigma_t)^{-(\lambda_2+1)}}{(1-\sigma_t)^{-\lambda_2}-1},\\
&\dfrac{\partial \mu_t}{\partial \eta_{1t}} = {\rm exp}(\eta_{1t})(1+\lambda_1{\rm exp}(\eta_{1t}))^{\frac{-(1+\lambda_1)}{\lambda_1}},\\
&\dfrac{\partial \sigma_t}{\partial \eta_{2t}} = {\rm exp}(\eta_{2t})(1+\lambda_2{\rm exp}(\eta_{2t}))^{\frac{-(1+\lambda_2)}{\lambda_2}},\\
&\rho_t=\dfrac{\partial\mu_t}{\partial\lambda_1} 
= \dfrac{1}{\lambda_1}\left[ \dfrac{1}{({\rm exp}(-\eta_{1t})+\lambda_1)} - \dfrac{{\rm log}(1+\lambda_1{\rm exp}(\eta_{1t}))}{\lambda_1} \right]  (1+\lambda_1 {\rm exp}(\eta_{1t}))^{-\frac{1}{\lambda_1}},
\end{align*}
and 
\begin{align*}
\varrho_t=\dfrac{\partial\sigma_t}{\partial\lambda_2} 
= \dfrac{1}{\lambda_2}\left[ \dfrac{1}{({\rm exp}(-\eta_{2t})+\lambda_2)} - \dfrac{{\rm log}(1+\lambda_2{\rm exp}(\eta_{2t}))}{\lambda_2} \right]  (1+\lambda_2 {\rm exp}(\eta_{2t}))^{-\frac{1}{\lambda_2}}.
\end{align*}
From these quantities, we can obtain the score vector and Fisher's information matrix given in Section \ref{s:emv}. These quantities assume that $\mu$ depends on $\lambda_1$ and $\sigma$ depends on $\lambda_2$.

\section{Numerical evaluation} \label{s:simu}

To assess the finite sample performance of the point estimators, this section provides a numerical evaluation using Monte Carlo simulations. This assessment considers the mean, bias, relative bias (RB), standard deviation (SD), and mean squared error (MSE) of the point estimates. We used $R=50,000$ Monte Carlo replications in each scenario, and considered sample sizes of $n=100$ and $n=500$. For each Monte Carlo replication, $n$ instances of the random variable $Y_t$ were generated with the density function in \eqref{E:densidadenova}, 
where the mean and dispersion parameters are given by $\mu_t=g_1^{-1}(\eta_{1t},\lambda_1)$ and  $\sigma_t=g_2^{-1}(\eta_{2t},\lambda_2)$, respectively. 
As discussed in Section~\ref{S:AO}, 
we considered two families of Aranda-Ordaz link functions, 
namely:  
symmetric and asymmetric. 
The values of $\boldsymbol{\beta}$, $\boldsymbol{\gamma}$, $\lambda_1$, and $\lambda_2$ are listed in Tables \ref{T:sym} and \ref{T:asym}, respectively, along with the numerical results.

The covariates for the mean and dispersion submodels were generated from the uniform distribution $(0,1)$, and were considered to be constant for all Monte Carlo replications. Computational implementations were conducted using the {\tt R} language~\citep{R2012}. 
An {\tt R} function for fitting the proposed model with asymmetric Aranda-Ordaz link function, 
along with the diagnostic measures, is available at \url{http://www.ufsm.br/bayer/betareglink.zip}.

\begin{table}[]
\setlength{\tabcolsep}{1.2pt}
\tablesize
\caption{Monte Carlo simulation results of point estimation evaluation for symmetric Aranda-Ordaz link functions.} \label{T:sym}
\begin{center}
\begin{tabular}{lrrrrrrrr}
 \hline
 \multicolumn{9}{c}{Scenario 1}\\
 \hline
 &$\beta_0$ & $\beta_1$& $\beta_2$& $\gamma_0$ & $\gamma_1$ & $\gamma_2$& $\lambda_1$ & $\lambda_2$ \\
 \hline
parameters	& $	1.500	$ & $	-1.000	$ & $	-1.500	$ & $	-1.700	$ & $	1.000	$ & $	-2.000	$ & $	0.500	$ & $	0.500	$ \\
 \hline
 \multicolumn{9}{c}{$n=100$}\\
 \hline
mean	& $	1.502	$ & $	-1.001	$ & $	-1.504	$ & $	-1.786	$ & $	0.780	$ & $	-1.126	$ & $	0.301	$ & $	0.688	$ \\
bias	& $	0.002	$ & $	-0.001	$ & $	-0.004	$ & $	-0.086	$ & $	-0.220	$ & $	0.874	$ & $	-0.199	$ & $	0.188	$ \\
RB	& $	0.151	$ & $	0.059	$ & $	0.251	$ & $	5.012	$ & $	-22.036	$ & $	-43.721	$ & $	-39.765	$ & $	37.686	$ \\
SD	& $	0.102	$ & $	0.185	$ & $	0.105	$ & $	0.190	$ & $	0.331	$ & $	0.553$ & $	0.308	$ & $	0.206	$ \\
MSE	& $	0.010	$ & $	0.034	$ & $	0.011	$ & $	0.043	$ & $	0.158	$ & $	1.070	$ & $	0.135	$ & $	0.078	$ \\
 \hline
 \multicolumn{9}{c}{$n=500$}\\
 \hline
mean	& $	1.508$ & $	-1.006	$ & $	-1.509	$ & $	-1.769	$ & $	0.871	$ & $	-1.376	$ & $	0.411	$ & $	0.626	$ \\
bias	& $	0.008	$ & $	-0.006	$ & $	-0.009	$ & $	-0.069	$ & $	-0.129	$ & $	0.626	$ & $	-0.089	$ & $	0.126	$ \\
RB	& $	0.509	$ & $	0.557	$ & $	0.612	$ & $	4.048	$ & $	-12.851	$ & $	-31.308	$ & $	-17.859	$ & $	25.229	$ \\
SD	& $	0.018	$ & $	0.017	$ & $	0.022	$ & $	0.146	$ & $	0.291	$ & $	0.623	$ & $	0.161	$ & $	0.219	$ \\
MSE	& $	0.000	$ & $	0.000	$ & $	0.001	$ & $	0.026	$ & $	0.101	$ & $	0.780	$ & $	0.034	$ & $	0.064	$ \\
\hline
 \multicolumn{9}{c}{Scenario 2}\\
 \hline
 &$\beta_0$ & $\beta_1$& $\beta_2$& $\gamma_0$ & $\gamma_1$ & $\gamma_2$& $\lambda_1$ & $\lambda_2$ \\
 \hline
parameters	& $	1.500	$ & $	-2.000	$ & $	1.000	$ & $	-2.000	$ & $	1.000	$ & $	-1.000	$ & $	0.250	$ & $	0.850$ \\
 \hline
 \multicolumn{9}{c}{$n=100$}\\
 \hline
mean	& $	1.507	$ & $	-2.008	$ & $	1.000	$ & $	-2.264	$ & $	0.699	$ & $	-0.713	$ & $	0.200	$ & $	0.614	$ \\
bias	& $	0.008	$ & $	-0.008	$ & $	-0.000	$ & $	-0.264	$ & $	-0.301	$ & $	0.287	$ & $	-0.050	$ & $	-0.237	$ \\
RB	& $	0.472	$ & $	0.389	$ & $	-0.027	$ & $	13.190	$ & $	-30.084	$ & $	-28.740	$ & $	-19.878	$ & $	-27.750	$ \\
SD	& $	0.053	$ & $	0.058	$ & $	0.048	$ & $	0.325	$ & $	0.494	$ & $	0.552	$ & $	0.125	$ & $	0.235	$ \\
MSE	& $	0.003	$ & $	0.003	$ & $	0.002	$ & $	0.175	$ & $	0.334	$ & $	0.387	$ & $	0.018	$ & $	0.111	$ \\
 \hline
 \multicolumn{9}{c}{$n=500$}\\
 \hline
mean	& $	1.503	$ & $	-2.004	$ & $	1.002	$ & $	-2.337	$ & $	0.880	$ & $	-0.876	$ & $	0.224	$ & $	0.612	$ \\
bias	& $	0.003	$ & $	-0.004	$ & $	0.002	$ & $	-0.337	$ & $	-0.120	$ & $	0.124	$ & $	-0.025	$ & $	-0.238	$ \\
RB	& $	0.171	$ & $	0.219	$ & $	0.231	$ & $	16.859	$ & $	-11.964	$ & $	-12.367	$ & $	-10.191	$ & $	-28.007	$ \\
SD	& $	0.015$ & $	0.026	$ & $	0.018$ & $	0.268	$ & $	0.496	$ & $	0.630	$ & $	0.079	$ & $	0.214	$ \\
MSE	& $	0.000	$ & $	0.001	$ & $	0.000	$ & $	0.185	$ & $	0.261	$ & $	0.412	$ & $	0.007	$ & $	0.102	$ \\
\hline
\end{tabular}
\end{center}
\end{table}

\begin{table}[]
\setlength{\tabcolsep}{1.2pt}
\tablesize
\caption{Monte Carlo simulation results of point estimation evaluation for asymmetric Aranda-Ordaz link functions.} \label{T:asym}
\begin{center}
\begin{tabular}{lrrrrrrrr}
 \hline
 \multicolumn{9}{c}{Scenario 1}\\
 \hline
 &$\beta_0$ & $\beta_1$& $\beta_2$& $\gamma_0$ & $\gamma_1$ & $\gamma_2$& $\lambda_1$ & $\lambda_2$ \\
 \hline
parameters	& $	1.000	$ & $	6.000	$ & $	-4.000	$ & $	-1.000	$ & $	-5.000	$ & $	3.000	$ & $	5.000	$ & $	10.000	$ \\
 \hline
 \multicolumn{9}{c}{$n=100$}\\
 \hline
mean	& $	1.019	$ & $	6.029	$ & $	-4.017	$ & $	0.415	$ & $	-7.122	$ & $	4.056	$ & $	5.025	$ & $	20.697	$ \\
bias	& $	0.019	$ & $	0.029	$ & $	-0.017	$ & $	1.415	$ & $	-2.122	$ & $	1.056	$ & $	0.025	$ & $	10.697	$ \\
RB	& $	1.090	$ & $	0.478	$ & $	0.434	$ & $	-141.521	$ & $	42.444	$ & $	35.207	$ & $	0.509	$ & $	106.971	$ \\
SD	& $	0.159	$ & $	0.415	$ & $	0.258	$ & $	5.459	$ & $	7.111	$ & $	3.392	$ & $	0.423	$ & $	36.193	$ \\
MSE	& $	0.0256	$ & $	0.173	$ & $	0.067	$ & $	31.805	$ & $	55.068	$ & $	12.618	$ & $	0.180	$ & $	1424.330	$ \\
 \hline
 \multicolumn{9}{c}{$n=500$}\\
 \hline
 mean	& $	1.007	$ & $	6.002	$ & $	-4.001	$ & $	-0.918	$ & $	-5.166	$ & $	3.103	$ & $	5.001	$ & $	10.829	$ \\
bias	& $	0.001	$ & $	0.002	$ & $	-0.001	$ & $	0.082	$ & $	-0.166	$ & $	0.103	$ & $	0.001	$ & $	0.829	$ \\
RB	& $	0.069	$ & $	0.025	$ & $	0.025	$ & $	-8.150	$ & $	3.330	$ & $	3.437	$ & $	0.029	$ & $	8.290	$ \\
SD	& $	0.046	$ & $	0.108	$ & $	0.072	$ & $	0.299	$ & $	0.442	$ & $	0.279	$ & $	0.112	$ & $	2.632	$ \\
MSE	& $	0.002	$ & $	0.012	$ & $	0.005	$ & $	0.096	$ & $	0.223	$ & $	0.089	$ & $	0.012	$ & $	7.616	$ \\

\hline
 \multicolumn{9}{c}{Scenario 2}\\
 \hline
 &$\beta_0$ & $\beta_1$& $\beta_2$& $\gamma_0$ & $\gamma_1$ & $\gamma_2$& $\lambda_1$ & $\lambda_2$ \\
 \hline
 parameters	& $	1.000	$ & $	3.000	$ & $	-4.000	$ & $	-1.000	$ & $	-8.000	$ & $	1.000	$ & $	1.000	$ & $	1.000	$ \\
 \hline
 \multicolumn{9}{c}{$n=100$}\\
 \hline
mean	& $	1.000	$ & $	3.000	$ & $	-4.000	$ & $	-0.863	$ & $	-8.301	$ & $	1.032	$ & $	1.000	$ & $	2.449$ \\
bias	& $	-0.000	$ & $	-0.000	$ & $	0.000	$ & $	0.137	$ & $	-0.301	$ & $	0.032	$ & $	-0.000	$ & $	1.449	$ \\
RB	& $	-0.001	$ & $	-0.000	$ & $	-0.000	$ & $	-13.685	$ & $	3.766	$ & $	3.241	$ & $	-0.001	$ & $	144.888	$ \\
SD	& $	0.002	$ & $	0.003	$ & $	0.003	$ & $	0.336	$ & $	0.491	$ & $	0.295	$ & $	0.002	$ & $	2.931	$ \\
MSE	& $	0.000	$ & $	0.000	$ & $	0.000	$ & $	0.136	$ & $	0.332	$ & $	0.088	$ & $	0.000	$ & $	10.692	$ \\
 \hline
 \multicolumn{9}{c}{$n=500$ }\\
 \hline
mean	& $	1.000	$ & $	3.000	$ & $	-4.000	$ & $	-0.976	$ & $	-8.059	$ & $	1.010	$ & $	1.000	$ & $	1.270	$ \\
bias	& $	-0.000	$ & $	-0.000	$ & $	0.000	$ & $	0.024	$ & $	-0.059	$ & $	0.010	$ & $	-0.000	$ & $	0.270	$ \\
RB	& $	-0.001	$ & $	-0.000	$ & $	-0.000	$ & $	-2.425	$ & $	0.736	$ & $	0.962	$ & $	-0.001	$ & $	26.986859	$ \\
SD	& $	0.001	$ & $	0.001	$ & $	0.001	$ & $	0.127	$ & $	0.185	$ & $	0.120	$ & $	0.001	$ & $	1.016	$ \\
MSE	& $	0.000	$ & $	0.000	$ & $	0.000	$ & $	0.017	$ & $	0.038	$ & $	0.015	$ & $	0.000	$ & $	1.105	$ \\
\hline
\end{tabular}
\end{center}
\end{table}

In general, 
according to 
Tables~\ref{T:sym} and \ref{T:asym}, 
the parameter estimates related to the mean submodel are not biased, unlike those for the dispersion submodel. This bias in the dispersion parameter estimators has been verified in other variations of the beta regression model, that consider fixed links \citep{Ospina2006,simas2010,Ospina2012}.
Considering symmetric family of Aranda-Ordaz link function, 
Table~\ref{T:sym} shows that the estimators for the dispersion submodel parameters  
are more biased than when we consider the asymmetric family (Table~\ref{T:asym}). 
We also note that the estimator of $\lambda_2$ 
was biased even in moderate sample sizes. 
This results can be justified by numerical problems in the log-likelihood maximization. 
The symmetric Aranda-Ordaz link function is numerically 
more unstable than the asymmetric one, 
due to fact that it fails to be differentiable at some points for some values of $\lambda$ (cf. Figure~\ref{F:ao-sym}).

For results about asymmetric family in Table~\ref{T:asym},
it can be observed that the bias in the dispersion structure estimators is concentrated at the intercept and for higher values of $\lambda_2$. 
The estimator of the link function parameter in the dispersion submodel also produced a considerable value of RB 
in small samples. 
For example, 
in Scenario 1, 

with $n=100$ and $\lambda_2=10$,  
${\rm RB}=-141.521\%$ for the intercept of the dispersion submodel.
As for $\lambda_2=2$, considering $n=100$, ${\rm RB}=106.971\%$ was observed for $\widehat{\lambda}_2$. 
This bias considerably decreases as the sample size increases; for $n=500$, the bias for the same estimators are reduced to $-8.150\%$ and $8.290\%$, respectively. In all cases, it is possible to verify that the MSE values tend fastly toward zero as the size of the sample increases, as was expected because of the consistency of the  MLEs.

The simulation results indicate that the MLE in the proposed model performs well. 
The bias in the dispersion submodel parameter estimators 
is in accordance with previews results \citep{Ospina2006, Andrade2007, simas2010, Ospina2012}. 
However, 
when the symmetric link was considered,  
the numerical maximization of the log-likelihood function presented some drawbacks. 
In addition, 
the asymmetric family of Aranda-Ordaz link function is more flexible than the symmetric version, 
because it considers the 
possible asymmetry between the random component and the linear predictors. 
This way, 
 we suggest the asymmetric family 
to empirical applications.

It is noteworthy that adequate link functions 
must be selected when using the usual models with fixed link functions (logit, probit, etc.) 
in actual data applications, in addition to the selection of the covariates. 
This model selection procedure can be time-consuming and inconclusive. 
When considering the proposed model, the selection of link functions is no longer a practical problem. 
Furthermore, the possible relationships between the parameters of interest, $\mu$ and $\sigma$, and their respective linear predictors, become more flexible.

\section{Application}\label{s:apli}

In this section, the proposed model is employed with actual data to demonstrate its practical applicability. 
For parametric link functions  
we choose the asymmetric Aranda-Ordaz family, 
because it is much more flexible than the symmetric function 
and its computational implementation is more stable.
We considered the data used by \cite{Cribari2013} about religious belief in 124 countries. 
The proportion of nonbelievers in each country is the dependent variable, $Y$. The covariates considered are the average intelligence quotient of the population in each country ($IQ$), $IQ$ squared ($IQ^2$), a dummy variable that equals 1 if the percentage of Muslims is greater than $50\%$ and 0 otherwise ($MUSL$), the per capita income adjusted by the purchasing power parity in 2008 in thousands of dollars ($INCOME$), the logarithm of the ratio between the sum of imports and exports and the Gross Domestic Product in 2008 (${\rm log}OPEN$), and the interaction between $MUSL$ and $INCOME$ ($M\times I$). 

After some adjustments and diagnostic analyses, the model presented in Table~\ref{t:par} was selected. The RESET-type test considering the LR  statistic suggests this model was correctly specified ($p\text{-value}=0.153$). It can also be verified that all covariates were significant at the nominal $10\%$ level. Comparatively, using the usual logit link function for the mean and dispersion, with the fitted model covariates given in Table~\ref{t:par}, the RESET-type test indicated that the model was not correctly specified ($p\text{-value}=0.008 $) at the usual nominal levels. 
We also tested the hypothesis $H_0: (\lambda_1, \lambda_2) = (1,1)$ by LR statistic. 
With $p\text{-value}=0.024$ we reject the hypothesis that the logit is the correct link function in both submodels.

\begin{table}[t]
\tablesize
\caption{Fitted model for religious belief data.} \label{t:par}
\begin{center}
\begin{tabular}{lcccc}
\hline
& Estimate & Std. error & $z$ stat & $p\text{-value}$ \\
\hline
\multicolumn{5}{c}{Mean submodel}\\
\hline
Intercept & $25.183$  &   $7.041$ & $3.576$ &  $0.000$\\
$IQ$     &   $-0.881$ &    $0.190$ & $4.623$ &  $0.000$\\
$IQ^2$   &   $0.006$ &    $0.001$ & $4.861$ &  $0.000$\\
$INCOME$ &   $0.029$ &    $0.017$ & $1.690$ &  $0.091$\\
$MUSL$   &   $-0.761$ &    $0.142$ & $5.354$ &  $0.000$\\
${\rm log}OPEN$  &   $0.481$  &   $0.162$ & $2.967$ &  $0.003$\\
$\lambda_1$ &    $9.255$ &  $3.892$  & &\\
\hline
\multicolumn{5}{c}{Dispersion submodel}\\
\hline
Intercept & $-8.817$  &   $1.354$ & $6.510$  & $0.000$\\
$IQ$        &  $0.059$  &   $0.011$ & $5.250$  & $0.000$\\
$MUSL$       &   $-1.608$  &   $0.256$ & $6.281$ &  $0.000$\\
${\rm log}OPEN$ &     $0.548$  &   $0.213$ & $2.580$ &  $0.010$\\
$M\times I$    &        $0.118$  &   $0.036$ & $3.308$ &  $0.001$\\
$\lambda_2$ &    $0.853$  &   $1.605$  & &\\
\hline
\multicolumn{5}{c}{$R^2_{G}=0.841$}\\
\multicolumn{5}{c}{${\rm AIC}=-560.271$.}\\
\hline
\end{tabular}
\end{center}
\end{table}

\begin{figure*}[p]
\label{F:f2}
\begin{center}
\subfigure[Standardized weighted residual 2.]{
{\includegraphics[width=0.45\textwidth]{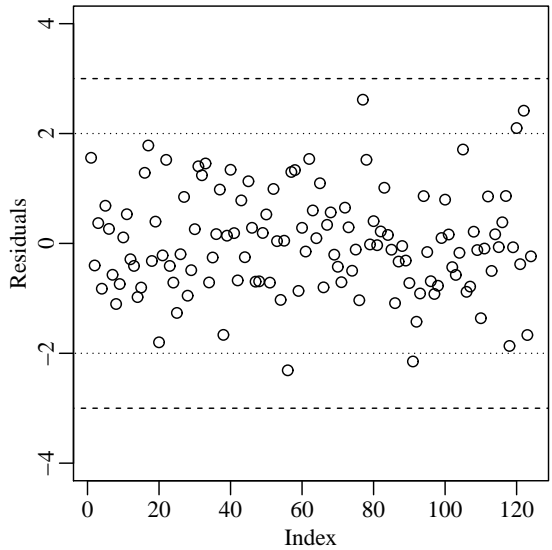}}\label{F:f2.2}}
\quad 
\subfigure[Fitted values versus observed values.]{
{\includegraphics[width=0.45\textwidth]{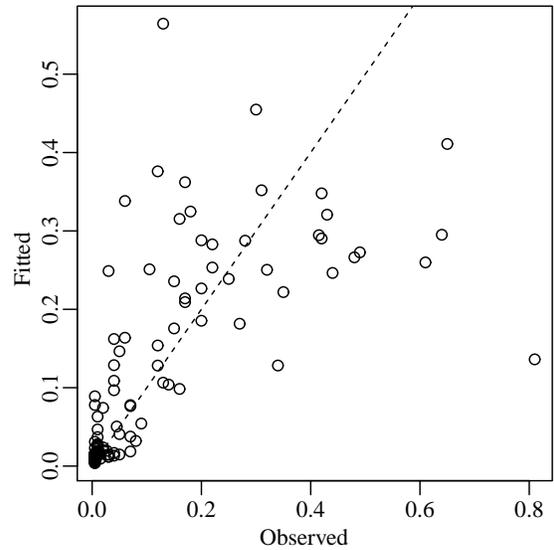}}\label{F:f2.3}}
\quad 
\subfigure[Half-normal probability plot.]{
{\includegraphics[width=0.45\textwidth]{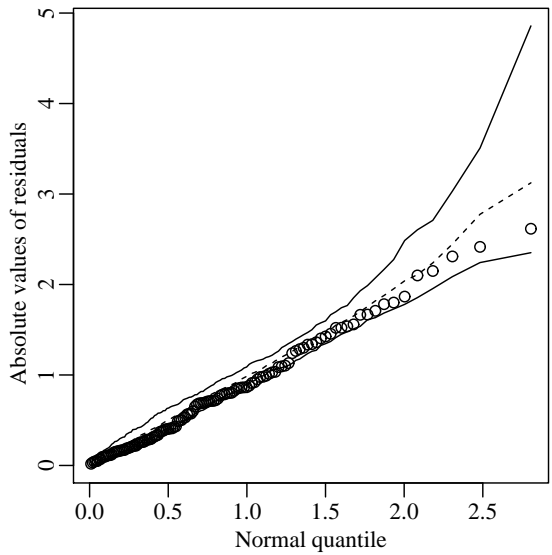}}\label{F:f2.4}}
\quad 
\subfigure[Cook-like distance.]{
{\includegraphics[width=0.45\textwidth]{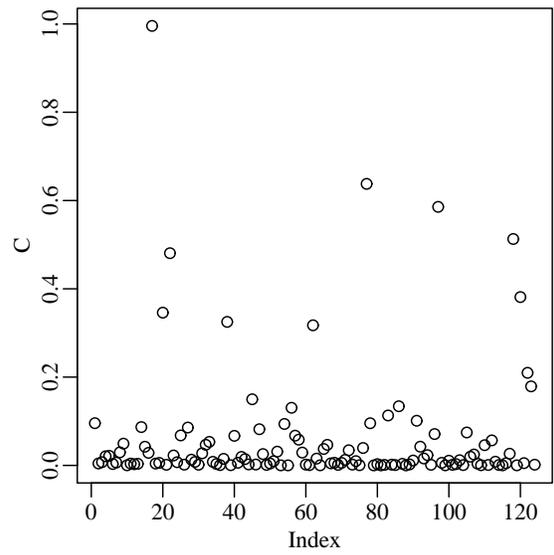}}\label{F:f2.5}}
\caption{Diagnostic charts.}
\label{F:f2}
\end{center}
\end{figure*}

Figure \ref{F:f2} presents a diagnostic analysis of the fitted model. The residual analysis in Figures \ref{F:f2.2} and \ref{F:f2.4}, and the observed values ($y_t$) versus the predicted values ($\widehat{\mu}_t$) in Figure \ref{F:f2.3}, indicates that the model was correctly adjusted. The Cook-like distance shown in Figure \ref{F:f2.5}, highlights four observations ($C_t>0.5$), namely: 17, 77, 97, and 118, corresponding to Burkina Faso, Mozambique, Sierra Leone, and the United States of America (USA), respectively. In Burkina Faso and Sierra Leone, just $0.5\%$ of the population are atheists, which is the smallest percentage of nonbelievers. 
Mozambique and Sierra Leone present the smallest average $IQ$ among the considered countries. In addition, Mozambique has a large proportion of atheists compared to other countries with similar $IQ$. Finally, the USA has very high $IQ$ and $INCOME$ values, as well as small $OPEN$ values compared with countries that present a similar percentage of nonbelievers (just $10.5\%$). Although the influence measures described by \cite{Cribari2013} did not highlight the USA, the authors did discuss this atypical religious characteristic for a country with high $IQ$.

\begin{figure*}[h]
\label{F:f3}
\begin{center}
\subfigure[$IQ$ versus estimated impact.]{
{\includegraphics[width=0.45\textwidth]{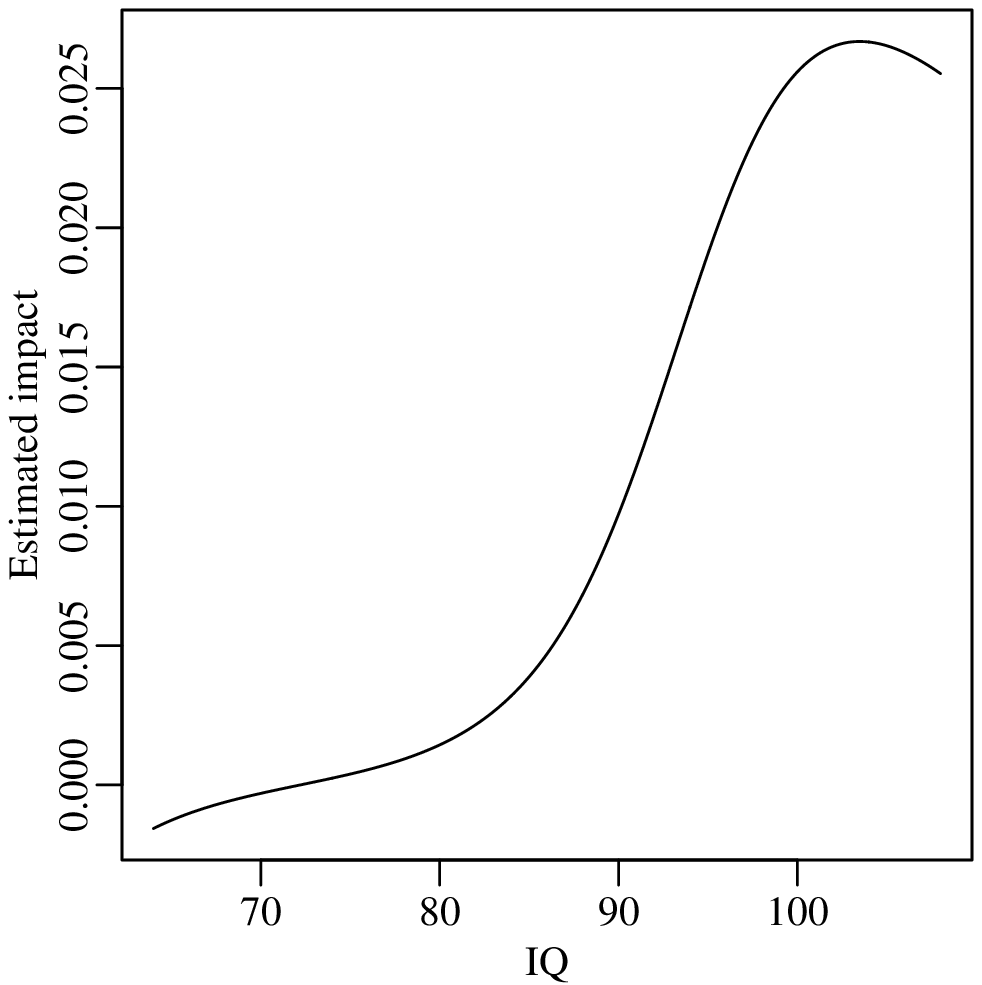}}\label{F:f3.1}}
\quad 
\subfigure[$IQ$ versus $\widehat{\mu}$.]{
{\includegraphics[width=0.45\textwidth]{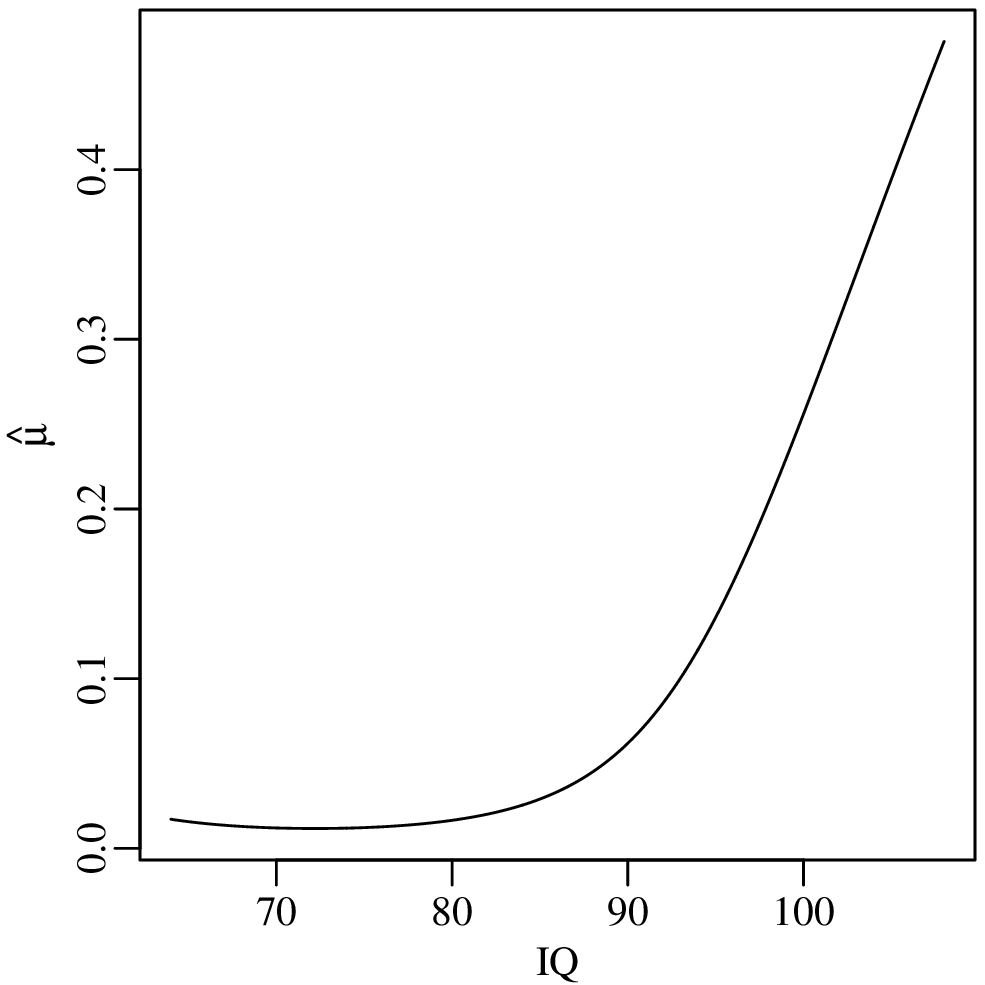}}\label{F:f3.2}}
\caption{Relationship between religious disbelief and intelligence.}
\end{center}
\end{figure*}

Conclusions regarding the mean submodel parameter estimates (Table \ref{t:par}) corroborate those of \cite{Cribari2013}. 
The variables $IQ$ and $MUSL$ have a negative influence on the mean submodel, whereas $IQ^2$, $INCOME$ and ${\rm log}OPEN$ have a positive influence. 
In the dispersion submodel, the variable $MUSL$ has a negative influence, whereas the variables $IQ$, ${\rm log}OPEN$, and $M\times I$  have a positive influence.
It is easy to see that the per capita income adjusted by the purchasing power parity ($INCOME$) is directly proportional to religious disbelief. 
To assess the impact of $IQ$ on the mean proportion of nonbelievers, the following measure of impact was considered \citep{Cribari2013}:
\begin{align*} 
\dfrac{\partial \Bbb{E}(y_t)}{\partial IQ_t} = \dfrac{\partial g_1^{-1}(\eta_{1t},\lambda_1)}{\partial IQ_t} = \dfrac{\partial \mu_t}{\partial IQ_t} = \dfrac{\partial \mu_t}{\partial \eta_{1t}} \dfrac{\partial \eta_{1t}}{\partial IQ_t}.
\end{align*}
This average impact on the proportion of nonbelievers resulting from changes in the $IQ$ covariate when the other covariates remain constant. Figure \ref{F:f3.1} shows the impact of variations in $IQ$ on the average percentage of nonbelievers, with the other covariates set to their mean values. The impact is not constant and varies according to $IQ$. Up to $IQ=100$, the impact first increases before decreasing. Figure \ref{F:f3.2} shows the relationship between the estimated mean proportion of nonbelievers and intelligence. This chart suggests that higher values of $IQ$ are related to larger proportions of nonbelievers, with greater impact for $IQ$ values above 85.

In order to compare our proposed model adjusted 
for religious belief data 
with 
the model 
in \cite{Cribari2013}, 
we elected some goodness-of-fit measures. 
The generalized coefficient of determination ($R^2_{G}$), 
the maximized log-likelihood function ($\ell(\widehat{\boldsymbol{\theta}})$), 
the Akaike information criterion (AIC) 
and 
the mean square error (MSE) between the observed ($y$) and predicted ($\widehat{\mu}$) values 
of the two fitted models are in Table~\ref{t:compare}. 
We note that our proposed model outperforms the model with fixed link functions in all measures. 
In particular, 
regarding $R^2_{G}$,  
our fitted model explains the variability of $y$ about $8\%$ more than the model with fixed links. 

It is worth noting that the proposed model considers the dispersion parameter $\sigma$, unlike the model used by \cite{Cribari2013}, which considered the precision parameter $\phi$. Note that \cite{Cribari2013} selected the loglog link function for the mean and the log link function for the precision.

\begin{table}
\tablesize
\caption{
A comparison between the proposed fitted model for religious belief data and the model in \cite{Cribari2013}.
} \label{t:compare}
\begin{center}
\begin{tabular}{ccccc}
\hline
Model & $R^2_{G}$ & $\ell(\widehat{\boldsymbol{\theta}})$ & AIC & MSE($y$,$\widehat{\mu}$) \\
\hline
Model with fixed links &
\multirow{2}{*}{$0.760$} & \multirow{2}{*}{$267.489$} & \multirow{2}{*}{$-518.979$} & \multirow{2}{*}{$0.015$}
\\
\citep{Cribari2013}
&  &  &  &  \\
\hline
Model with parametric links  & \multirow{2}{*}{$0.841$} & \multirow{2}{*}{$293.135$} & \multirow{2}{*}{$-560.271$} & \multirow{2}{*}{$0.013$} 
\\
(proposed)
&  &  &  &  \\
\hline
\end{tabular}
\end{center}
\end{table}

\section{Conclusion} \label{s:conclu}

In this paper, we have proposed a beta regression model with parametric link functions, that is useful for modeling variables contained in the interval $(0,1)$, such as rates and proportions. The vector score and Fisher's information matrix were derived analytically, and aspects of large sample inference  were presented. Diagnostic measures that allow researchers to identify influential points, outlier observations, or shortcomings of the fitted model were also proposed. A simulation study highlighted the accurate finite sample performance of the point estimators. An application to actual data was presented and discussed to demonstrate the practical usefulness of the proposed model. Moreover, the use of parametric link functions enables problems arising from the incorrect specification of link functions to be circumvented, thereby facilitating the construction of an adequate model. Finally, all of the evidence from this study suggests that the proposed model is both useful and adequate for modeling rate and proportion variables.

\section*{Acknowledgements}

This research was partially supported by 
Conselho Nacional de Desenvolvimento Científico e Tecnológico (CNPq), Brazil. 
The final publication is available at Springer via http://dx.doi.org/10.1007/s00362-017-0885-9.

\section*{Appendix}\label{S:apA}

In this appendix we obtain the score function and the Fisher's information matrix
for all parameters ($\boldsymbol{\beta}$,$\boldsymbol{\gamma}$,$\lambda_1$,$\lambda_2$). 

The elements of the score vector are given by:
\begin{align*}
U_{\beta_i}(\boldsymbol{\theta})=\frac{\partial\ell(\boldsymbol{\theta})}{\partial\beta_i}=&\sum\limits_{t=1}^{n}\dfrac{\partial\ell_{t}(\mu_{t},\sigma_{t})}{\partial\mu_t}\frac{\partial\mu_t}{\partial\eta_{1t}}\frac{\partial\eta_{1t}}{\partial\beta_{i}},\\
U_{\gamma_j}(\boldsymbol{\theta})=\frac{\partial\ell(\boldsymbol{\theta})}{\partial\gamma_j}=&\sum_{t=1}^{n}\dfrac{\partial\ell_{t}(\mu_{t},\sigma_{t})}{\partial\sigma_t}\dfrac{\partial\sigma_t}{\partial\eta_{2t}}\dfrac{\partial\eta_{2t}}{\partial\gamma_{j}},\\
U_{\lambda_1}(\boldsymbol{\theta})=\frac{\partial\ell(\boldsymbol{\theta})}{\partial\lambda_1} =& \sum_{t=1}^{n}\dfrac{\partial\ell_{t}(\mu_{t},\sigma_{t})}{\partial\mu_t}\dfrac{\partial\mu_t}{\partial\lambda_1},\\
U_{\lambda_2}(\boldsymbol{\theta})=\frac{\partial\ell(\boldsymbol{\theta})}{\partial\lambda_2} =& \sum_{t=1}^{n}\dfrac{\partial\ell_{t}(\mu_{t},\sigma_{t})}{\partial\sigma_t}\dfrac{\partial\sigma_t}{\partial\lambda_2},
\end{align*}
for $i=1,\ldots,r$ and $j=1, \ldots, s$, 
where $\dfrac{\partial\ell_{t}(\mu_{t},\sigma_{t})}{\partial\mu_t} = \dfrac{1-\sigma^2_t}{\sigma^2_t}(y^*_t-\mu^*_t)$, $\dfrac{\partial\mu_t}{\partial\eta_{1t}} = \left[\dfrac{\partial g_1(\mu_{t},\lambda_1)}{\partial\mu_t}\right]^{-1}$, $\dfrac{\partial\eta_{1t}}{\partial\beta_{i}}=x_{ti}$, $\dfrac{\partial\ell_{t}(\mu_{t},\sigma_{t})}{\partial\sigma_t}=a_t$, $\dfrac{\partial\sigma_t}{\partial\eta_{2t}} = \left[\dfrac{\partial g_2(\sigma_{t},\lambda_2)}{\partial\sigma_t}\right]^{-1}$ and 
$\dfrac{\partial\eta_{2t}}{\partial\gamma_{i}}=z_{tj}$.

The second order derivatives of the log-likelihood function are given by:
\begin{align*}
\dfrac{\partial^2\ell(\boldsymbol{\theta})}{\partial\beta_i\partial\beta_p} 
&= \sum_{t=1}^{n}\dfrac{\partial}{\partial\mu_t}
\left( \dfrac{\partial\ell_t(\mu_t,\sigma_t)}{\partial\mu_t}\dfrac{\partial\mu_t}{\partial\eta_{1t}} \right)  \dfrac{\partial\mu_t}{\partial\eta_{1t}}  \dfrac{\partial\eta_{1t}}{\partial\beta_p}   \dfrac{\partial\eta_{1t}}{\partial\beta_i} \\
&= \sum_{t=1}^{n} \left( \dfrac{\partial_2\ell_t(\mu_t,\sigma_t)}{\partial\mu_t^2}\dfrac{\partial\mu_t}{\partial\eta_{1t}} + \dfrac{\partial\ell_t(\mu_t,\sigma_t)}{\partial\mu_t} \dfrac{\partial}{\partial\mu_t} \left(\dfrac{\partial\mu_t}{\partial\eta_{1t}}\right) \right) \\
 &\times \left( \dfrac{\partial g_1(\mu_t,\lambda_1)}{\partial\mu_t} \right)^{-1} x_{ti}x_{tp}, \; p=1,\ldots,r,\\
\dfrac{\partial^2\ell(\boldsymbol{\theta})}{\partial\beta_i\partial\gamma_j} &= \sum_{t=1}^{n} 
\left(\dfrac{\partial^2\ell_{t}(\mu_{t},\sigma_{t})}{\partial\mu_t\partial\sigma_t}\dfrac{\partial\sigma_t}{\partial\eta_{2t}}\dfrac{\partial\eta_{2t}}{\partial\gamma_{j}} \right)\frac{\partial\mu_t}{\partial\eta_{1t}} \frac{\partial\eta_{1t}}{\partial\beta_{i}} \\
&=\sum_{t=1}^{n} 
\left(\dfrac{\partial^2\ell_{t}(\mu_{t},\sigma_{t})}{\partial\mu_t\partial\sigma_t}\left( \dfrac{\partial g_2(\sigma_t,\lambda_2)}{\partial\sigma_t} \right)^{-1} z_{tj} \right) \left(\dfrac{\partial g_1(\mu_t,\lambda_1)}{\partial\mu_t} \right)^{-1} x_{ti},\\
\dfrac{\partial^2\ell(\boldsymbol{\theta})}{\partial\beta_i\partial\lambda_1} &= \sum_{t=1}^{n} 
\dfrac{\partial}{\partial\lambda_1} \left(\dfrac{\partial\ell_{t}(\mu_{t},\sigma_{t})}{\partial\mu_t}\dfrac{\partial\mu_t}{\partial\eta_{1t}} \right)\frac{\partial\eta_{1t}}{\partial\beta_{i}}\\
&=\sum_{t=1}^{n} \bigg[ \dfrac{\partial^2\ell_{t}(\mu_{t},\sigma_{t})}{\partial\mu_t^2} \dfrac{\partial\mu_t}{\partial\lambda_1} \dfrac{\partial\mu_t}{\partial\eta_{1t}} + \dfrac{\partial\ell_{t}(\mu_{t},\sigma_{t})}{\partial\mu_t} \dfrac{\partial}{\partial\lambda_1} \left( \dfrac{\partial\mu_t}{\partial\eta_{1t}}  \right) \bigg] x_{ti},\\
\dfrac{\partial^2\ell(\boldsymbol{\theta})}{\partial\beta_i\partial\lambda_2} &= \sum_{t=1}^{n} 
\dfrac{\partial}{\partial\lambda_2} \left(\dfrac{\partial\ell_{t}(\mu_{t},\sigma_{t})}{\partial\mu_t}\dfrac{\partial\mu_t}{\partial\eta_{1t}} \right)\frac{\partial\eta_{1t}}{\partial\beta_{i}}\\
&=\sum_{t=1}^{n} \bigg[ \dfrac{\partial^2\ell_{t}(\mu_{t},\sigma_{t})}{\partial\mu_t\partial\sigma_t} \dfrac{\partial\sigma_t}{\partial\lambda_2} \dfrac{\partial\mu_t}{\partial\eta_{1t}} + \dfrac{\partial\ell_{t}(\mu_{t},\sigma_{t})}{\partial\mu_t} \dfrac{\partial}{\partial\lambda_2} \left( \dfrac{\partial\mu_t}{\partial\eta_{1t}}  \right) \bigg] x_{ti}\\
& = \sum_{t=1}^{n}  \dfrac{\partial^2\ell_{t}(\mu_{t},\sigma_{t})}{\partial\mu_t\partial\sigma_t} \varrho_t  \left(\dfrac{\partial g_1(\mu_t,\lambda_1)}{\partial\mu_t} \right)^{-1} x_{ti},\\
\dfrac{\partial^2\ell(\boldsymbol{\theta})}{\partial\gamma_j\partial\gamma_l} &= \sum_{t=1}^{n} \dfrac{\partial}{\partial\gamma_l}\left( \dfrac{\partial\ell_{t}(\mu_{t},\sigma_{t})}{\partial\sigma_t}\dfrac{\partial\sigma_t}{\partial\eta_{2t}}\dfrac{\partial\eta_{2t}}{\partial\gamma_{j}} \right) \\
&= \sum_{t=1}^{n} \bigg( \dfrac{\partial^2\ell_{t}(\mu_{t},\sigma_{t})}{\partial\sigma_t^2} \dfrac{\partial\sigma_t}{\partial\eta_{2t}}\dfrac{\partial\eta_{2t}}{\partial\gamma_{j}} \dfrac{\partial\sigma_t}{\partial\eta_{2t}} + \dfrac{\partial\ell_{t}(\mu_{t},\sigma_{t})}{\partial\sigma_t} \dfrac{\partial}{\partial\gamma_l} \left( \dfrac{\partial\sigma_t}{\partial\eta_{2t}} \right) \bigg)z_{tj},\\
l=1,\ldots,s,\\
\dfrac{\partial^2\ell(\boldsymbol{\theta})}{\partial\gamma_j\partial\lambda_1} &= \sum_{t=1}^{n} \dfrac{\partial}{\partial\lambda_1}\left( \dfrac{\partial\ell_{t}(\mu_{t},\sigma_{t})}{\partial\sigma_t}\dfrac{\partial\sigma_t}{\partial\eta_{2t}}\dfrac{\partial\eta_{2t}}{\partial\gamma_{j}} \right)\\
& = \sum_{t=1}^{n}  \dfrac{\partial^2\ell_{t}(\mu_{t},\sigma_{t})}{\partial\sigma_t\partial\mu_t}\dfrac{\partial\mu_t}{\partial\lambda_1}  \left( \dfrac{\partial g_2(\sigma_t,\lambda_2)}{\partial\sigma_t} \right)^{-1}  z_{tj},\\
\dfrac{\partial^2\ell(\boldsymbol{\theta})}{\partial\gamma_j\partial\lambda_2} &= \sum_{t=1}^{n} \dfrac{\partial}{\partial\lambda_2}\left( \dfrac{\partial\ell_{t}(\mu_{t},\sigma_{t})}{\partial\sigma_t}\dfrac{\partial\sigma_t}{\partial\eta_{2t}}\dfrac{\partial\eta_{2t}}{\partial\gamma_{j}} \right)\\ 
&= \sum_{t=1}^{n} \bigg( \dfrac{\partial^2\ell_{t}(\mu_{t},\sigma_{t})}{\partial\sigma_t^2}\dfrac{\partial\sigma_t}{\partial\lambda_2} \dfrac{\partial\sigma_t}{\partial\eta_{2t}} + \dfrac{\partial\ell_{t}(\mu_{t},\sigma_{t})}{\partial\sigma_t} \dfrac{\partial}{\partial\lambda_2}\left( \dfrac{\partial\sigma_t}{\partial\eta_{2t}} \right) \bigg) z_{tj},\\
\dfrac{\partial^2\ell(\boldsymbol{\theta})}{\partial\lambda_1^2} &= \sum_{t=1}^{n} \dfrac{\partial}{\partial\lambda_1}\left( \dfrac{\partial\ell_{t}(\mu_{t},\sigma_{t})}{\partial\mu_t}\dfrac{\partial\mu_t}{\partial\lambda_1}\right) \\
&= \sum_{t=1}^{n} \bigg( \dfrac{\partial^2\ell_{t}(\mu_{t},\sigma_{t})}{\partial\mu_t^2}\dfrac{\partial\mu_t}{\partial\lambda_1} \dfrac{\partial\mu_t}{\partial\lambda_1}  \dfrac{\partial\ell_{t}(\mu_{t},\sigma_{t})}{\partial\mu_t} \dfrac{\partial^2\mu}{\partial\lambda_1^2}  \bigg),\\
\dfrac{\partial^2\ell(\boldsymbol{\theta})}{\partial\lambda_1\partial\lambda_2} &= \sum_{t=1}^{n} \dfrac{\partial}{\partial\lambda_2}\left( \dfrac{\partial\ell_{t}(\mu_{t},\sigma_{t})}{\partial\mu_t}\dfrac{\partial\mu_t}{\partial\lambda_1}\right) = \sum_{t=1}^{n} \dfrac{\partial^2\ell_{t}(\mu_{t},\sigma_{t})}{\partial\mu_t\partial\sigma_t}\dfrac{\partial\sigma_t}{\partial\lambda_2} \dfrac{\partial\mu_t}{\partial\lambda_1},\\
\dfrac{\partial^2\ell(\boldsymbol{\theta})}{\partial\lambda_2^2} &= \sum_{t=1}^{n} \dfrac{\partial}{\partial\lambda_2}\left( \dfrac{\partial\ell_{t}(\mu_{t},\sigma_{t})}{\partial\sigma_t}\dfrac{\partial\sigma_t}{\partial\lambda_2}\right) \\
&= \sum_{t=1}^{n} \bigg( \dfrac{\partial^2\ell_{t}(\mu_{t},\sigma_{t})}{\partial\sigma_t^2}\dfrac{\partial\sigma_t}{\partial\lambda_2} \dfrac{\partial\sigma_t}{\partial\lambda_2}  \dfrac{\partial\ell_{t}(\mu_{t},\sigma_{t})}{\partial\sigma_t} \dfrac{\partial^2\sigma}{\partial\lambda_2^2}  \bigg),
\end{align*}
where $\dfrac{\partial}{\partial\lambda_2} \left( \dfrac{\partial\mu_t}{\partial\eta_{1t}}  \right)=0$, 
\begin{align*}
\dfrac{\partial^2\ell_t(\mu_t,\sigma_t)}{\partial\mu_t^2} &= - \left(\dfrac{1-\sigma_t^2}{\sigma_t^2}\right)^2 \bigg[ \psi'\left( \mu_t \dfrac{1-\sigma_t^2}{\sigma_t^2}  \right) + \psi'\left( (1-\mu_t) \dfrac{1-\sigma_t^2}{\sigma_t^2}  \right)  \bigg],\\
\dfrac{\partial^2\ell_{t}(\mu_{t},\sigma_{t})}{\partial\mu_t\partial\sigma_t}&=-\dfrac{2}{\sigma_t^3}(y^*_t-\mu^*_t) - \dfrac{1-\sigma^2_t}{\sigma^2_t}\dfrac{2}{\sigma_t^3} \bigg[ (1-\mu_t)\psi'\left( (1-\mu_t) \dfrac{1-\sigma_t^2}{\sigma_t^2}  \right) \\
&-\mu_t\psi'\left( \mu_t \dfrac{1-\sigma_t^2}{\sigma_t^2}  \right)   \bigg],\\
\dfrac{\partial^2\ell_{t}(\mu_{t},\sigma_{t})}{\partial\sigma_t^2}
&= -\dfrac{4}{\sigma^6_t}\bigg[-\psi'\left(\dfrac{1-\sigma_t^2}{\sigma_t^2}\right) + \mu_t^2\psi'\left(\mu_t\dfrac{1-\sigma_t^2}{\sigma_t^2}\right) + (1-\mu_t)^2\\
&\times\psi'\left((1-\mu_t)\dfrac{1-\sigma_t^2}{\sigma_t^2}\right)\bigg] +\dfrac{3}{\sigma_t}\dfrac{2}{\sigma_t^3} \bigg[ \mu_t(y_t^*-\mu_t^*)+\psi\left(\frac{1-\sigma_t^2}{\sigma_t^2}\right) \\
&-\psi\left((1-\mu_t)\frac{1-\sigma_t^2}{\sigma_t^2}\right)+\log (1-y_t)  \bigg].
\end{align*}

Taking the expected value of the second order derivatives given above, 
since $\Bbb{E}\left( \dfrac{\partial\ell_t(\mu_t,\sigma_t)}{\partial\mu_t} \right) = 0$, 
we have:
\begin{align*}
\Bbb{E}\left( \dfrac{\partial\ell(\boldsymbol{\theta})}{\partial\beta_i\partial\beta_p} \right) &= \sum_{t=1}^{n} \Bbb{E} \Bigg[ \left( \dfrac{\partial_2\ell_t(\mu_t,\sigma_t)}{\partial\mu_t^2}\left( \dfrac{\partial g_1(\mu_t,\lambda_1)}{\partial\mu_t} \right)^{-1} \right)  \left( \dfrac{\partial g_1(\mu_t,\lambda_1)}{\partial\mu_t} \right)^{-1}\!\!\!\! x_{ti}x_{tp} \Bigg]\\
&= \sum_{t=1}^{n} \Bbb{E} \left[ \dfrac{\partial_2\ell_t(\mu_t,\sigma_t)}{\partial\mu_t^2}\left( \dfrac{\partial g_1(\mu_t,\lambda_1)}{\partial\mu_t} \right)^{-2}  x_{ti}x_{tp} \right]\\
&= - \sum_{t=1}^{n} \Bbb{E} \Bigg[ \left(\dfrac{1-\sigma_t^2}{\sigma_t^2}\right) \left(\dfrac{1-\sigma_t^2}{\sigma_t^2}\right) \bigg[ \psi'\left( \mu_t \dfrac{1-\sigma_t^2}{\sigma_t^2}  \right) \\
&+  \psi'\left( (1-\mu_t) \dfrac{1-\sigma_t^2}{\sigma_t^2}  \right)  \bigg]\left( \dfrac{\partial g_1(\mu_t,\lambda_1)}{\partial\mu_t} \right)^{-2}  x_{ti}x_{tp} \Bigg]\\
&= - \sum_{t=1}^{n} \left(\dfrac{1-\sigma_t^2}{\sigma_t^2}\right) w_t x_{ti} x_{tp}.
\end{align*}
Since 
\begin{align*}
\Bbb{E}\left(\dfrac{\partial^2\ell_{t}(\mu_{t},\sigma_{t})}{\partial\mu_t\partial\sigma_t}\right)&=- \dfrac{1-\sigma^2_t}{\sigma^2_t}\dfrac{2}{\sigma_t^3}\bigg[ (1-\mu_t)\psi'\bigg( (1-\mu_t) \dfrac{1-\sigma_t^2}{\sigma_t^2}  \bigg)\\
&-\mu_t\psi'\left( \mu_t \dfrac{1-\sigma_t^2}{\sigma_t^2}  \right) \bigg],
\end{align*}
we arrive at the conclusion that
\begin{align*}
\Bbb{E}\left(\dfrac{\partial^2\ell(\boldsymbol{\theta})}{\partial\beta_i\partial\gamma_j}\right)&=-\sum_{t=1}^{n}c_t \left( \dfrac{\partial g_2(\sigma_t,\lambda_2)}{\partial\sigma_t} \right)^{-1}  \left(\dfrac{\partial g_1(\mu_t,\lambda_1)}{\partial\mu_t} \right)^{-1} z_{tj} x_{ti}.
\end{align*}
In relation to $\beta_i$ and $\lambda_1$, we have:
\begin{align*}
\Bbb{E}\left(\dfrac{\partial^2\ell(\boldsymbol{\theta})}{\partial\beta_i\partial\lambda_1}\right) &= \sum_{t=1}^{n} \Bbb{E}\left( \dfrac{\partial^2\ell_{t}(\mu_{t},\sigma_{t})}{\partial\mu_t^2} \dfrac{\partial\mu_t}{\partial\lambda_1} \dfrac{\partial\mu_t}{\partial\eta_{1t}} \right) = \sum_{t=1}^{n}  \dfrac{\partial^2\ell_{t}(\mu_{t},\sigma_{t})}{\partial\mu_t^2} \dfrac{\partial\mu_t}{\partial\lambda_1} \dfrac{\partial\mu_t}{\partial\eta_{1t}} \\
&= -\sum_{t=1}^{n} \nu_t \rho_t \left(\dfrac{\partial g_1(\mu_t,\lambda_1)}{\partial\mu_t} \right)^{-1} x_{ti}.
\end{align*}

The expected value of the second order derivative with respect to  $\beta_i$ and $\lambda_2$ is given by:
\begin{align*}
\Bbb{E}\left(\dfrac{\partial^2\ell(\boldsymbol{\theta})}{\partial\beta_i\partial\lambda_2}\right)\! &= \sum_{t=1}^{n}  \Bbb{E}\left(\!\dfrac{\partial^2\ell_{t}(\mu_{t},\sigma_{t})}{\partial\mu_t\partial\sigma_t}\right)\! \varrho_t \!\left(\! \dfrac{\partial g_1(\mu_t,\lambda_1)}{\partial\mu_t} \right)^{\!\!-1} \!\!\!\!x_{ti} \\
&= \sum_{t=1}^{n}  c_t \varrho_t  \left(\dfrac{\partial g_1(\mu_t,\lambda_1)}{\partial\mu_t} \right)^{-1} x_{ti}.
\end{align*}

Since $\Bbb{E}\left(\dfrac{\partial\ell_{t}(\mu_{t},\sigma_{t})}{\partial\sigma_t}\right)=0$,  
we have
\begin{align*}
&\Bbb{E}\left(\dfrac{\partial^2\ell(\boldsymbol{\theta})}{\partial\gamma_j\partial\gamma_l}\right) 
= \sum_{t=1}^{n} \Bbb{E}\left(\dfrac{\partial^2\ell_{t}(\mu_{t},\sigma_{t})}{\partial\sigma_t^2}\right) \!\!\left( \frac{g_2(\sigma_t,\lambda_2)}{\partial\sigma_t}  \right)^{\!\!-2}\!\!\!\! z_{tl} z_{tj},
\end{align*}
where
\begin{align*}
\Bbb{E}\left(\dfrac{\partial^2\ell_{t}(\mu_{t},\sigma_{t})}{\partial\sigma_t^2}\right) 
 &= -\dfrac{4}{\sigma^6_t}\bigg[-\psi'\left(\dfrac{1-\sigma_t^2}{\sigma_t^2}\right) 
 + \mu_t^2\psi'\left(\mu_t\dfrac{1-\sigma_t^2}{\sigma_t^2}\right) \\
&+ (1-\mu_t)^2\psi'\left((1-\mu_t)\dfrac{1-\sigma_t^2}{\sigma_t^2}\right)\bigg].
\end{align*}

With respect to $\gamma_j$ and $\lambda_1$, we have:
\begin{align*}
\Bbb{E}\left(\dfrac{\partial^2\ell(\boldsymbol{\theta})}{\partial\gamma_j\partial\lambda_1}\right)
\! &= \sum_{t=1}^{n}  \Bbb{E}\left(\! \dfrac{\partial^2\ell_{t}(\mu_{t},\sigma_{t})}{\partial\sigma_t\partial\mu_t}  \right) \! \rho_t \!\left(\! \dfrac{\partial g_2(\sigma_t,\lambda_2)}{\partial\sigma_t} \right)^{\!\!-1} \!\!\!\! z_{tj} \\
&= - \sum_{t=1}^{n}  c_t  \rho_t \left( \dfrac{\partial g_2(\sigma_t,\lambda_2)}{\partial\sigma_t} \right)^{-1}  z_{tj}.
\end{align*}

For $\gamma_j$ and $\lambda_2$, we have:
\begin{align*}
\Bbb{E}\left(\dfrac{\partial^2\ell(\boldsymbol{\theta})}{\partial\gamma_j\partial\lambda_2}\right)\! &= \sum_{t=1}^{n}  \Bbb{E}\left( \!\dfrac{\partial^2\ell_{t}(\mu_{t},\sigma_{t})}{\partial\sigma_t^2}  \right) \! \varrho_t \!\left( \!\dfrac{\partial g_2(\sigma_t,\lambda_2)}{\partial\sigma_t} \right)^{\!-1} \!\!\!\! z_{tj} \\
&= - \sum_{t=1}^{n}  d_t^*  \varrho_t \left( \dfrac{\partial g_2(\sigma_t,\lambda_2)}{\partial\sigma_t} \right)^{-1}  z_{tj}.
\end{align*}

Finally, we have: 
\begin{align*}
\Bbb{E}\left(\dfrac{\partial^2\ell(\boldsymbol{\theta})}{\partial\lambda_1^2}\right) = -\sum_{t=1}^{n}\nu_t\rho_t\rho_t,
\end{align*}
\begin{align*}
&\Bbb{E}\left(\dfrac{\partial^2\ell(\boldsymbol{\theta})}{\partial\lambda_1\partial\lambda_2}\right)\! = \sum_{t=1}^{n} \Bbb{E}\left(\!\dfrac{\partial^2\ell_t(\mu_t\sigma_t)}{\partial\mu_t\partial\sigma_t}\right)\! \varrho_t\rho_t=\! -\!\sum_{t=1}^{n} c_t\varrho_t\rho_t,
\end{align*}
and
\begin{align*}
\Bbb{E}\left(\dfrac{\partial^2\ell(\boldsymbol{\theta})}{\partial\lambda_2^2}\right)\! =\sum_{t=1}^{n} \Bbb{E}\left( \!\dfrac{\partial^2\ell_t(\mu_t,\sigma_t)}{\partial\sigma_t^2} \right) \! \varrho_t\varrho_t  = \!-\!\sum_{t=1}^{n}d_t^*\varrho_t\varrho_t.
\end{align*}

In matrix form, we have:
\begin{align*}
\Bbb{E}\left( \dfrac{\partial\ell(\boldsymbol{\theta})}{\partial\beta_i\partial\beta_p} \right) &= -\boldsymbol{X}^\top \boldsymbol{\Sigma} \boldsymbol{W}\boldsymbol{X},\\
\Bbb{E}\left(\dfrac{\partial^2\ell(\boldsymbol{\theta})}{\partial\beta_i\partial\gamma_j}\right)&=-\boldsymbol{X}^\top \boldsymbol{C}\boldsymbol{T}\boldsymbol{H}\boldsymbol{Z},\\
\Bbb{E}\left(\dfrac{\partial^2\ell(\boldsymbol{\theta})}{\partial\beta_i\partial\lambda_1}\right) &= -\boldsymbol{X}^\top \boldsymbol{V}\boldsymbol{T}\boldsymbol{\rho},\\
\Bbb{E}\left(\dfrac{\partial^2\ell(\boldsymbol{\theta})}{\partial\beta_i\partial\lambda_2}\right) &= -\boldsymbol{X}^\top \boldsymbol{C}\boldsymbol{T}\boldsymbol{\varrho},\\
\Bbb{E}\left(\dfrac{\partial^2\ell(\boldsymbol{\theta})}{\partial\gamma_j\partial\gamma_l}\right) &= -\boldsymbol{Z}^\top \boldsymbol{D}^*\boldsymbol{H}\boldsymbol{H}^\top \boldsymbol{Z},\\
\Bbb{E}\left(\dfrac{\partial^2\ell(\boldsymbol{\theta})}{\partial\gamma_j\partial\lambda_1}\right) &= -\boldsymbol{Z}^\top \boldsymbol{C}\boldsymbol{H}\boldsymbol{\rho},\\
\Bbb{E}\left(\dfrac{\partial^2\ell(\boldsymbol{\theta})}{\partial\gamma_j\partial\lambda_2}\right) &= -\boldsymbol{Z}^\top \boldsymbol{D}^*\boldsymbol{H}\boldsymbol{\varrho},\\
\Bbb{E}\left(\dfrac{\partial^2\ell(\boldsymbol{\theta})}{\partial\lambda_1^2}\right) &= -\boldsymbol{\rho}^\top \boldsymbol{V}\boldsymbol{\rho},\\
\Bbb{E}\left(\dfrac{\partial^2\ell(\boldsymbol{\theta})}{\partial\lambda_1\partial\lambda_2}\right) &= -\boldsymbol{\rho}^\top \boldsymbol{C}\boldsymbol{\varrho},\\
\Bbb{E}\left(\dfrac{\partial^2\ell(\boldsymbol{\theta})}{\partial\lambda_2^2}\right) &= -\boldsymbol{\varrho}^\top \boldsymbol{D}^*\boldsymbol{\varrho}.
\end{align*}


\begin{thebibliography}{58}
\providecommand{\natexlab}[1]{#1}
\providecommand{\url}[1]{{#1}}
\providecommand{\urlprefix}{URL }
\expandafter\ifx\csname urlstyle\endcsname\relax
  \providecommand{\doi}[1]{DOI~\discretionary{}{}{}#1}\else
  \providecommand{\doi}{DOI~\discretionary{}{}{}\begingroup
  \urlstyle{rm}\Url}\fi
\providecommand{\eprint}[2][]{\url{#2}}

\bibitem[{Adewale and Xu(2010)}]{Adewale2010}
Adewale AJ, Xu X (2010) Robust designs for generalized linear models with
  possible overdispersion and misspecified link functions. Computational
  Statistics \& Data Analysis 54(4):875--890

\bibitem[{Akaike(1974)}]{Akaike1974}
Akaike H (1974) A new look at the statistical model identification. IEEE
  Transactions on Automatic Control 19(6):716--726

\bibitem[{Akaike(1983)}]{Akaike1983}
Akaike H (1983) Information measures and model selection. Bulletin of the
  International Statistical Institute 50:277--290

\bibitem[{Andrade(2007)}]{Andrade2007}
Andrade ACG (2007) Efeitos da especifica\c{c}\~ao incorreta da fun\c{c}\~ao de
  liga\c{c}\~ao no modelo de regress\~ao beta. Master's thesis, Universidade
  Federal de S\~ao Paulo

\bibitem[{Aranda-Ordaz(1981)}]{Ordaz1981}
Aranda-Ordaz FJ (1981) On two families of transformations to additivity for
  binary response data. Biometrika 68(2):357--363

\bibitem[{Atkinson(1981)}]{Atkinson1981}
Atkinson A (1981) Two graphical display for outlying and influential
  observations in regression. Biometrika 68(1):13--20

\bibitem[{Atkinson(1985)}]{Atkinson1985}
Atkinson AC (1985) Plots, Transformations and Regression: An Introduction to
  Graphical Methods of Diagnostic Regression Analysis. New York: Oxford
  University Press

\bibitem[{Bayer and Cribari-Neto(2017)}]{Bayer2015}
Bayer FM, Cribari-Neto F (2017) Model selection criteria in beta regression
  with varying dispersion. Communications in Statistics - Simulation and
  Computation 46(1):729--746

\bibitem[{Colosimo et~al(2000)Colosimo, Chalita, and Dem\'etrio}]{Colosimo2000}
Colosimo EA, Chalita LVAS, Dem\'etrio CGB (2000) Tests of proportional
  {H}azards and proportional odds models for grouped survival data. Biometrics
  56(4):1233--1240

\bibitem[{Cook(1977)}]{Cook1977}
Cook RD (1977) Detection of influential observations in linear regression.
  Technometrics 19(1):15--18

\bibitem[{Cox and Reid(1987)}]{Cox1987}
Cox DR, Reid N (1987) Parameter orthogonality and approximate conditional
  inference. Journal of the Royal Statistical Society Series B 49(1):1--39

\bibitem[{Cribari-Neto and Souza(2012)}]{Souza2012}
Cribari-Neto F, Souza TC (2012) Testing inference in variable dispersion beta
  regressions. Journal of Statistical Computation and Simulation
  82(12):1827--1843

\bibitem[{Cribari-Neto and Souza(2013)}]{Cribari2013}
Cribari-Neto F, Souza TC (2013) Religious belief and intelligence: Worldwide
  evidence. Intelligence 41(5):482--489

\bibitem[{Czado(1994)}]{Czado1994}
Czado C (1994) Parametric link modification of both tails in binary regression.
  Statistical Papers 35(1):189--201

\bibitem[{Czado(1997)}]{Czado1997}
Czado C (1997) On selecting parametric link transformation families in
  generalized linear models. Journal of Statistical Planning and Inference
  61(1):125--139

\bibitem[{Czado and Raftery(2006)}]{Czado2006}
Czado C, Raftery AE (2006) Choosing the link function and accounting for link
  uncertainty in generalized linear models using {B}ayes factors. Statistical
  Papers 47(3):419--442

\bibitem[{Dehbi et~al(2014)Dehbi, Cortina-Borja, and Geraci}]{Dehbi2014}
Dehbi H, Cortina-Borja M, Geraci M (2014) AOfamilies: Aranda-Ordaz
  Transformation Families.
  \urlprefix\url{http://cran.r-project.org/package=AOfamilies}, {R} Package

\bibitem[{Dehbi et~al(2016)Dehbi, Cortina-Borja, and Geraci}]{Dehbi2016}
Dehbi HM, Cortina-Borja M, Geraci M (2016) {A}randa-{O}rdaz quantile regression
  for student performance assessment. Journal of Applied Statistics 43(1):58--71 

\bibitem[{Espinheira et~al(2008{\natexlab{a}})Espinheira, Ferrari, and
  Cribari-Neto}]{espinheira2008b}
Espinheira P, Ferrari SLP, Cribari-Neto F (2008{\natexlab{a}}) On beta
  regression residuals. Journal of Applied Statistics 35(4):407--419

\bibitem[{Espinheira et~al(2008{\natexlab{b}})Espinheira, Ferrari, and
  Cribari-Neto}]{espinheira2008}
Espinheira PL, Ferrari SLP, Cribari-Neto F (2008{\natexlab{b}}) Influence
  diagnostics in beta regression. Computational Statistics \& Data Analysis
  52(9):4417--4431

\bibitem[{Ferrari and Cribari-Neto(2004)}]{Ferrari2004}
Ferrari SLP, Cribari-Neto F (2004) Beta regression for modelling rates and
  proportions. Journal of Applied Statistics 31(7):799--815

\bibitem[{Ferrari and Pinheiro(2011)}]{Pinheiro2011}
Ferrari SLP, Pinheiro EC (2011) Improved likelihood inference in beta
  regression. Journal of Statistical Computation and Simulation 81(4):431--443

\bibitem[{Ferrari et~al(2011)Ferrari, Espinheira, and
  Cribari-Neto}]{Ferrari2011}
Ferrari SLP, Espinheira PL, Cribari-Neto F (2011) Diagnostic tools in beta
  regression with varying dispersion. Statistica Neerlandica 65(3):337--351

\bibitem[{Geraci and Jones(2015)}]{Geraci2015}
Geraci M, Jones MC (2015) Improved transformation-based quantile regression.
  The Canadian Journal of Statistics 43(1):118--132

\bibitem[{Gomes and Ludermir(2013)}]{Gomes2013}
Gomes GSdS, Ludermir TB (2013) Optimization of the weights and asymmetric
  activation function family of neural network for time series forecasting.
  Expert Systems with Applications 40(16):6438--6446

\bibitem[{Guerrero and Johnson(1982)}]{Guerrero1982}
Guerrero VM, Johnson RA (1982) Use of the {B}ox-{C}ox transformation with
  binary response models. Biometrika 69(2):309--314

\bibitem[{Kaiser(1997)}]{Kaiser1997}
Kaiser MS (1997) Maximum likelihood estimation of link function parameters.
  Computational Statistics \& Data Analysis 24(1):79--87

\bibitem[{Koenker and Yoon(2009)}]{Koenker2009}
Koenker R, Yoon J (2009) Parametric links for binary choice models: A
  {F}isherian-{B}ayesian colloquy. Journal of Econometrics 152(2):120--130

\bibitem[{McCullagh and Nelder(1989)}]{McCullagh1989}
McCullagh P, Nelder J (1989) Generalized linear models, 2nd edn. Chapman and
  Hall

\bibitem[{Morgan(1992)}]{Morgan1992}
Morgan BJ (1992) Analysis of Quantal Response Data. Chapman and Hall/CRC

\bibitem[{Nagelkerke(1991)}]{Nagelkerke1991}
Nagelkerke NJD (1991) A note on a general definition of the coefficient of
  determination. Biometrika 78(3):691--692.

\bibitem[{Neyman and Pearson(1928)}]{Person1928}
Neyman J, Pearson ES (1928) On the use and interpretation of certain test
  criteria for purposes of statistical inference. Biometrika 20A(1/2):175--240

\bibitem[{Oliveira(2013)}]{Oliveira2013}
Oliveira JSC (2013) Detectando m\'a especifica\c{c}\~ao em regress\~ao beta.
  Master's thesis, Universidade Federal de Pernanbuco

\bibitem[{Ospina and Ferrari(2012)}]{Ospina2012}
Ospina R, Ferrari SLP (2012) A general class of zero-or-one inflated beta
  regression models. Computational Statistics \& Data Analysis 56(6):1609--1623

\bibitem[{Ospina et~al(2006)Ospina, Cribari-Neto, and
  Vasconcellos}]{Ospina2006}
Ospina R, Cribari-Neto F, Vasconcellos KLP (2006) Improved point and intervalar
  estimation for a beta regression model. Computational Statistics \& Data
  Analysis 51(2):960--981

\bibitem[{Paolino(2001)}]{Paolino2001}
Paolino P (2001) Maximum likelihood estimation of models with beta-distributed
  dependent variables. Political Analysis 9(4):325--346

\bibitem[{Pawitan(2001)}]{Pawitan2001}
Pawitan Y (2001) In All Likelihood: Statistical Modelling and Inference Using
  Likelihood. Oxford Science publications

\bibitem[{Pereira and Cribari-Neto(2013)}]{Pereira2013}
Pereira TL, Cribari-Neto F (2013) Detecting model misspecification in inflated
  beta regressions. Communications in Statistics - Simulation and Computation
  43(3):631--656

\bibitem[{Pregibon(1980)}]{Pregibon1980}
Pregibon D (1980) Goodness of link tests for generalized linear models. Applied
  Statistics 29(1):15--24

\bibitem[{Press et~al(1992)Press, Teukolsky, Vetterling, and Flannery}]{press}
Press W, Teukolsky S, Vetterling W, Flannery B (1992) Numerical recipes in {C}:
  {T}he art of scientific computing, 2nd edn. Cambridge University Press

\bibitem[{{R Development Core Team}(2014)}]{R2012}
{R Development Core Team} (2014) R: A Language and Environment for Statistical
  Computing. R Foundation for Statistical Computing, Vienna, Austria, {ISBN}
  3-900051-07-0

\bibitem[{Ramalho et~al(2011)Ramalho, Ramalho, and Murteira}]{Ramalho2011}
Ramalho EA, Ramalho JJ, Murteira JMR (2011) Alternative estimating and testing
  empirical strategies for fractional regression models. Journal of Economic
  Surveys 25(1):16--68

\bibitem[{Ramsey(1969)}]{Ramsey1969}
Ramsey JB (1969) Tests for specification errors in classical linear
  least-squares regression analysis. Journal of the Royal Statistical Society Series B 
  31(2):350--371

\bibitem[{Rao(1948)}]{Rao1948}
Rao C (1948) Large sample tests of statistical hypotheses concerning several
  parameters with applications to problems of estimation. Mathematical
  Proceedings of the Cambridge Philosophical Society 44(1):50--57

\bibitem[{Rigby and Stasinopoulos(2005)}]{Rigby2005}
Rigby R, Stasinopoulos D (2005) Generalized additive models for location, scale
  and shape (with discussion). Applied Statistics 54(3):507--554

\bibitem[{Scallan et~al(1984)Scallan, Guilchrist, and Green}]{Scallan1984}
Scallan A, Guilchrist R, Green M (1984) Fitting parametric link functions in
  generalized linear models. Computational Statistics \& Data Analysis
  2(1):37--49

\bibitem[{Schwarz(1978)}]{Schwarz1978}
Schwarz G (1978) Estimating the dimension of a model. Annals of Statistics
  6(2):461--464

\bibitem[{Simas et~al(2010)Simas, Barreto-Souza, and Rocha}]{simas2010}
Simas AB, Barreto-Souza W, Rocha AV (2010) Improved estimators for a general
  class of beta regression models. Computational Statistics \& Data Analysis
  54(2):348--366

\bibitem[{Smith(2003)}]{Smith2003}
Smith DM (2003) Computing single parameter transformations. Communications in
  Statistics - Simulation and Computation 32(3):605--618

\bibitem[{Smithson and Verkuilen(2006)}]{smithson2006}
Smithson M, Verkuilen J (2006) A better lemon squeezer? {M}aximum-likelihood
  regression with beta-distributed dependent variables. Psychol Methods
  11(1):54--71

\bibitem[{Smyth and Verbyla(1999)}]{Smyth1999}
Smyth GK, Verbyla AP (1999) Adjusted likelihood methods for modelling
  dispersion in generalized linear models. Environmetrics 10(6):695--709

\bibitem[{Stukel(1988)}]{Stukel1988}
Stukel TA (1988) Generalized logistic models. Journal of the American
  Statistical Association 83(402):426--431

\bibitem[{Taneichi et~al(2014)Taneichi, Sekiya, and Toyama}]{Taneichi2014}
Taneichi N, Sekiya Y, Toyama J (2014) A new family of parametric links for
  binomial generalized linear models. Journal of the Japan Statistical Society
  44(2):119--133

\bibitem[{Terrell(2002)}]{Terrell2002}
Terrell GR (2002) The gradient statistic. Computing Science and Statistics
  34:206--215

\bibitem[{Vargas et~al(2014)Vargas, Ferrari, and Lemonte}]{Vargas2014}
Vargas TM, Ferrari SL, Lemonte AJ (2014) Improved likelihood inference in
  generalized linear models. Computational Statistics \& Data Analysis
  74:110--124

\bibitem[{Wald(1943)}]{Wald1943}
Wald A (1943) Tests of statistical hypotheses concerning several parameters
  when the number of observations is large. Transactions of the American
  Mathematical Society 54:426--482

\bibitem[{Zhao et~al(2014)Zhao, Zhang, Lv, and Liu}]{Zhao2014}
Zhao W, Zhang R, Lv Y, Liu J (2014) Variable selection for varying dispersion
  beta regression model. Journal of Applied Statistics 41(1):95--108

\bibitem[{Zimprich(2010)}]{Zimprich2010}
Zimprich D (2010) Modeling change in skewed variables using mixed beta
  regression models. Research in Human Development 7(1):9--26

\end{thebibliography}

\end{document}